\documentclass[aps,showpacs,onecolumn,preprint,superscriptaddress]{revtex4-1}
\usepackage{
epsfig,amssymb,amsmath,amsfonts}

\usepackage{caption}
\usepackage{subcaption}
\usepackage{graphicx}
\usepackage{color}

\newcommand{\daga}[1]{\ensuremath{#1^{\dagger}}}

\begin{document}

\title{Quantum phase transitions of atom-molecule Bose mixtures in a double-well potential}

\author{A. Rela\~no}
\affiliation{Departamento de  F\'{\i}sica Aplicada I and GISC, Universidad Complutense de
  Madrid}
\author{J. Dukelsky}
\affiliation{Instituto de Estructura de la Materia, CSIC, Serrano 123,
  28006 Madrid}
\author{P. P\'erez-Fern\'andez}
\affiliation{Departamento de F\'{\i}sica Aplicada III, Universidad de
  Sevilla}
\author{J.M. Arias}
\affiliation{Departamento de F\'{\i}sica At\'omica, Molecular y
  Nuclear, Universidad de Sevilla and Unidad Asociada to IEM (CSIC) Madrid}

\date{\today}

\begin{abstract}
  The ground state and spectral properties of Bose gases in
  double-well potentials are studied in two different scenarios: i) an
  interacting atomic Bose gas, and ii) a mixture of an atomic gas
  interacting with diatomic molecules. A ground state second-order
  quantum phase transition (QPT) is observed in both scenarios. For
  large attractive values of the atom-atom interaction, the
  ground-state is degenerate. For repulsive and small attractive
  interaction, the ground-state is not degenerate and is well
  approximated by a boson coherent state. Both systems depict an
  excited state quantum phase transition (ESQPT). In both
    cases, a critical energy separates a region in which all the
    energy levels are degenerate in pairs, from another region in
    which there are no degeneracies. For the atomic system, the
    critical point displays a singularity in the density of states,
    whereas this behavior is largely smoothed for the mixed
    atom-molecule system.
\end{abstract}

\maketitle

\section{Introduction}
Bose-Einstein condensation in dilute atomic gases provides a valuable
tool to study fundamental problems of quantum systems at a macroscopic
scale. Among such phenomena, the quantum transition from Josephson
oscillations to self-trapping is one of the most interesting. It
comprises the transition from a regime in which two spatially
separated Bose-Einstein condensates (BEC) oscillate, to another one in
which the macroscopic wave function remains trapped in one of the two
trap sides.  This transition has been experimentally observed in
\cite{Albeiz:05} by means of a condensate of $^{87}\text{Rb}$ atoms
in an optical trap. It has also been theoretically studied in Refs.
\cite{Leggett:01, Links:05, Julia:10}. Using a two-site Bose-Hubbard
Hamiltonian as a description of two BECs that can tunnel between the
two wells, it has been shown that the quantum phase transition (QPT)
develops at a critical value of the atom-atom interaction. The
imbalance, that is, the difference of population between the two
wells, acts as an order parameter for the transition, changing from
zero in the Josephson oscillation regime to a non-zero value in the
self-trapping regime.

This is an example of QPT which has been
studied in other systems. In \cite{Konotop:09}, the same kind of
transition was reported in a non-linear version of the same system, and it was
related to a classical bifurcation in the Hamiltonian. In
\cite{Zibold:10}, a similar bifurcation was linked to a transition
from zero to non-zero values in the imbalance.

In this work, we expand these studies by focusing not only in the
ground state QPT, but also looking at quantum critical phenomena in
the energy spectrum. In particular, we will describe the so called
excited-state quantum phase transitions (ESQPTs). These occur when
some kind of non-analytic behavior \cite{Caprio:08} arise in an
excited region of the energy spectrum, normally a singularity in the
density of states or in the flow of the energy spectrum through the
critical energy line. In contrast to usual quantum phase transitions
\cite{Sachdev}, they are not characterized by a critical value of the
control parameter $\lambda_c$, but by a critical energy $E_c$ above the
ground state, at which the non-analyticity takes place. Usually,
ESQPTs are linked to QPTs; below the critical parameter $\lambda_c$ no
transition appears, while above $\lambda_c$ the non-analytic behavior
characteristic of the QPT propagates to a certain critical excitation
energy $E_c$. However, in some cases an ESQPT occurs without the
corresponding QPT. As a paradigmatic example, we mention the
realization of the Lipkin-Meshkov-Glick model studied in
\cite{Zibold:10}.

ESQPTs entail dramatic dynamic consequences, like the enhancement of
the decoherence if the ESQPT takes place in the environment with which
a quantum system interacts \cite{Relano:08}, the onset of chaos
\cite{Pedro:11b} or the emergence of symmetry-breaking equilibrium
states after a quench \cite{Puebla:13}. They appear in models
pertaining to different branches of Physics, like the interacting
boson model (IBM) \cite{Arias:03,Cejnar:10}, Lipkin-Meshkov-Glick
model \cite{Ribeiro:08}, vibron model \cite{Perez-Bernal:10}, Dicke
and Jaynes-Cummings models \cite{Pedro:11,Pedro:11b,Brandes:13},
kicked top \cite{Bastidas:13} and microwave Dirac billiards
\cite{Dietz:13}. It is worth mentioning that, despite the fact that
the excitation energy is linked to the temperature of an isolated
system, ESQPTs are qualitatively different from thermal phase
transitions. The existence of a critical energy does not necessarily
entail a corresponding critical temperature. For example, the Dicke
model restricted to the maximum angular momentum sector displays an
ESQPT, but has no thermal phase transition \cite{Aparicio:12}.

In order to investigate these issues, we treat two related
systems. The first consists of an atomic Bose gas trapped in a
two-well potential, and the second incorporates the interaction of the
atoms with a diatomic molecule. Mixed condensates of atoms and
molecules have been experimentally realized by means of the
manipulation of a Feshbach
resonance \cite{Timmermans:99,Jin:05,Santos:06}. Mixtures of atoms and
molecules in double-well potentials have been studied previously in
\cite{Motohashi:10}. In \cite{Sanders:11} a similar system was used to
model the atom-atom interaction by means of a Feshbach resonance.

The organization of the paper is as follows. We introduce the two
models in Sect. II. In Sect. III and Sect. IV the ground state quantum
phase transitions are investigated numerically as well as analytically
for both models; first, within a mean field approach, for which we
later incorporate fluctuations. Sect. V is devoted to analyze ESQPTs
in both models. Finally, in Sect. VI the main conclusions of this work
are summarized.

\section{The models}

We shall study a bosonic atom gas confined in a double-well potential
as a model of a bosonic Josephson junction \cite{Milburn:97}. It has
been recently shown that a restriction to the lowest two modes is a
valid approximation for a wide range of the bosonic interaction
\cite{Julia:10}. In this limit, the model reduces to the two-site
Bose-Hubbard Hamiltonian. Each site represents the lowest mode of the
right and left wells, respectively,
\begin{equation}
H_1=-J \left( \daga{a_L} a_R + \daga{a_R} a_L \right) + \frac{U}{N}
\left( \daga{a_L} \daga{a_L} a_L a_L + \daga{a_R} \daga{a_R} a_R a_R
\right),
\label{Two}
\end{equation}
where the $\daga{a_k}$ and $a_k$ ($k=L, \, R$ for left and right
wells) operators are the atom creation and annihilation operators in
each well and obey the usual boson commutation rules, $N$ is the total
number of atoms, $J$ is the hopping strength, and $U$ is atomic
interaction. Note that the scaling factor $1/N$ in front
  of the atomic interaction is needed to ensure that the energy per
  particle $E/N$ is well-defined in the thermodynamic limit. The
static and dynamic properties of the Hamiltonian (\ref{Two}) have been
extensively studied in recent years \cite{Leggett:2001, Hol:2001,
  Chuchem:2010, Mele:2011}. Moreover, it has been shown that
(\ref{Two}) is a quantum integrable model and a particular limit of
Richardson-Gaudin bosonic families
\cite{Duke:2001,Links:2003,Duke:2005}.

In addition to (\ref{Two}), we will consider here an extension of the
two-site Hubbard model by the explicit inclusion of an interacting
diatomic molecule. The resulting two-channel Hamiltonian is: 

\begin{eqnarray}
H_2&=&-J \left( \daga{a_L} a_R + \daga{a_R} a_L \right) + \frac{U}{N}
\left( \daga{a_L} \daga{a_L} a_L a_L + \daga{a_R} \daga{a_R} a_R a_R
\right) \nonumber \\
&+& \omega \daga{b} b - \frac{g}{\sqrt{2 N}} \left[ \daga{b}
  \left( a_L a_L + a_R a_R \right) + \left( \daga{a_L} \daga{a_L} +
  \daga{a_R} \daga{a_R} \right) b \right],
\label{eq:Hamiltonian}
\end{eqnarray}
where $N$ is the total number of atoms plus twice the number of
molecules, $N=2 b^{\dagger} b + \left( \daga{a_L} a_L + \daga{a_R} a_R
\right)$, and $\daga{b}(b)$ creates (annihilates) a molecule. The
operators $b$ and $\daga{b}$ follow the usual bosonic commutation
rules. Because of the shape of the atom-molecule interaction (two
atoms are destroyed to create a molecule, and viceversa) they
represent a dimeric entity, giving rise to the previously written
conservation of the total number of particles $N$. The frequency
$\omega$ represents the self-energy of the molecule. When it goes to
zero, the Hamiltonian (\ref{eq:Hamiltonian}) describes a Feshbach
resonance interacting with the atomic cloud. In what follows we will
use the hopping term $J$ as the energy unit. This Hamiltonian is in
general non-integrable. However, as a function of the interaction
parameters $\omega$, $U$, and $g$ it has two integrable limits which
will allow us to explore the transition from regularity to
chaos. As in the previous case, the scaling factors $1/N$
  and $1/\sqrt{N}$ in front of the atom-atom and the atom-molecule
  interaction ensure that the energy per particle is well-defined in
  the thermodynamic limit.

For both Hamiltonians, (\ref{Two}) and (\ref{eq:Hamiltonian}), $N$ is
a conserved quantity. They also conserve
a kind of parity, related to the interchange between atoms in left and
right wells. 

In connection with this symmetry, it is worth mentioning that the
population imbalance operator that counts the difference between the
number of atoms in the left and right wells
\begin{equation}
\hat{I}=\daga{a_R} a_R - \daga{a_L} a_L,
\end{equation}
does not commute with the parity operator. Therefore, its expectation
value is zero in any common eigenstate of the Hamiltonian and the
parity operator.

\section{Phase diagram}

In this section we analyze and compare the ground-state phase diagram
of both models, the double well with and without the interaction with
the diatomic molecule.  In order to carry out this comparison, we
select the atomic interaction $U$ that is common to both models as the
control parameter, and keep fixed $\omega$ and $g$.  As it will be
seen below, the ground state of one of the phases is doubly degenerate
due to spontaneous symmetry-breaking. Thus, the population imbalance
is a good candidate of order parameter for this transition.  In fact,
it has been used in \cite{Julia:10} to characterize the macroscopic
self-trapping in the two-site Bose-Hubbard model. However, it is not
possible to rely directly on the population imbalance to study
numerically this transition, because both Hamiltonians, (\ref{Two})
and (\ref{eq:Hamiltonian}), have a well defined parity and, therefore,
the imbalance expectation value is zero in any of their
eigenstates. One possible solution is to introduce a small
symmetry-breaking term to break the degeneracies, as it has been done
in \cite{Julia:10}. In this work we do not follow this
  way; we will use, instead, a new order parameter derived from a
quantum version of the Fisher information, defined as
\cite{Mazzarella:11}
\begin{equation}
F_{QFI} = \left< \left( \widehat{n}_R - \widehat{n}_L \right)^2
\right> - \left( \left< \widehat{n}_R - \widehat{n}_L \right>
\right)^2.
\end{equation}
For eigenstates with well-defined parities, this expression reduces to:
\begin{equation}
F_{QFI} = \left< \left( \widehat{n}_R - \widehat{n}_L \right)^2
\right>.
\end{equation}
Therefore, to obtain a quantity proportional to the number
  of left or right-well atoms, we will use $\sqrt{F_{QFI} }$ as the
  order parameter of the transition. As it will be seen below, this
  magnitude behaves in the same way as the population imbalance in
  the exact solution of both Hamiltonians.

\subsection{Mean-field approximation}

The mean-field or semi-classical approximation is expected to
provide accurate results for this kind of two-level systems in the
large $N$ limit. In fact, it has been shown that the ground state
energy in this approximation is exact to leading order in $N$
\cite{GilFeng78}. The great advantage of this approach is that it can
be solved analytically. We use boson coherent states as variational
wavefunctions to model the ground state of both Hamiltonians. The
corresponding wavefunctions are:
\begin{equation}
 |\Psi_1 \rangle =
e^{\sqrt{N}(\gamma_{R}a^{\dagger}_{R}+\gamma_{L}a^{
\dagger}_{L})}| 0 \rangle,
\end{equation}
for the case without molecule, and
\begin{equation}
 |\Psi_2 \rangle =
e^{\sqrt{N}(\frac{\beta}{2}b^{\dagger}+\gamma_{R}a^{\dagger}_{R}+\gamma_{L}a^{
\dagger}_{L})}| 0 \rangle,
\end{equation}
for the case with molecule. They are eigenstates of the annihilation
operators $a_L$, $a_R$ (both) and $b$ (only the last), with
eigenvalues $\sqrt{N} \gamma_R$, $\sqrt{N} \gamma_L$, and $\sqrt{N}
\beta /2$, respectively; hence, they fulfill the mathematical
requirement for coherent states. The former could be obtained from the
second one as the limit in which the atom-molecule coupling parameters
$g$ and the corresponding amplitude in the coherent state ($\beta$) go
to zero, and the self-energy of the molecule $\omega$ goes to
infinity: $g\rightarrow 0$, $\omega \rightarrow \infty$ and $\beta
\rightarrow 0$. Therefore we will mostly concentrate on the second
model, and treat the other one as a particular case in the appropriate
limit. It is important to note that both coherent states break the
parity symmetry and will give a finite population imbalance different
from zero in the symmetry-broken phase.

\bigskip

The energy per particle using the coherent state $|\Psi_2 \rangle$ is
\begin{equation}
  \frac{E}{N}= \frac{\langle \Psi_2|H_2|\Psi_2 \rangle}{N\langle\Psi_2
    |\Psi_2 \rangle}=  -2J\gamma_R \gamma_L +
  \frac{\omega}{2}\beta^2-g\beta(\gamma_R^2+\gamma_L^2)+U(\gamma_R^4+\gamma_L^4)
  ,
\end{equation}
where the variational parameters are assumed to be real, provided
$g>0$. By minimizing this energy surface for the corresponding set of
parameters $J$, $\omega$, $g$ and $U$, optimal variational parameters
$\beta, \gamma_R$ and $\gamma_L$ are obtained, providing an
approximation to the exact ground state wavefunction and energy. This
method is only exact, as mentioned above, for the leading order in $N$
\cite{GilFeng78}. Because of the conservation of the total number of
particles in the system, there is a constraint for the variational
parameters: $\beta^2+\gamma_R^2+\gamma_L^2=1$. This condition is
included in the minimization by using a Lagrange multiplier which
introduces a new parameter, $\lambda$.

When the energy surface, with the constraint of the particle number
conservation, is minimized, the optimal values of the variational
parameters as a function of the control parameters are determined from
the set of equations,
\begin{eqnarray}
\beta^2+\gamma^2_R+\gamma^2_L & = &1, \label{energysurface1} \\
-g(\gamma^2_R+\gamma^2_L)+\omega \beta -2\lambda \beta & = & 0, \\
g\beta \gamma_R-2U\gamma^3_R+J\gamma_L+\lambda \gamma_R & = & 0, \\
J\gamma_R+g\beta \gamma_L-2U\gamma^3_L+\lambda \gamma_L & = & 0.
\label{energysurface2}
\end{eqnarray}
This set of four non linear equations has two classes of solutions: a)
the symmetric, with $\gamma_R = \gamma_L = \gamma$; and the
non-symmetric, b) $\gamma_R \not = \gamma_L$. The former corresponds
to the symmetric phase; the latter, to the non-symmetric phase.

\bigskip

The solutions in the symmetric phase are
\begin{equation}
\beta=\left\{ \begin{array}{ll}
-\frac{g}{2U}+\frac{(1-i\sqrt{3})A}{6\sqrt[3]{4}U\left(B+\sqrt{B^2+4A^3}\right)^{1/3}}-\frac{(1+i\sqrt{3})\left(B+\sqrt{B^2+4
    A^3}\right)^{1/3}}{12\sqrt[3]{2}U}  &  \mbox{if }
U\in (U_c,0)  \\
& \\
\frac{-2J-\omega+\sqrt{12g^2+(2J+\omega)^2}}{6g} &  \mbox{if } U=0\\
& \\
-\frac{g}{2U}-\frac{A}{3\sqrt[3]{4}U\left(B+\sqrt{B^2+4
    A^3}\right)^{1/3}} +\frac{\left(B+\sqrt{B^2+4
    A^{3}}\right)^{1/3}}{6\sqrt[3]{2}U} &  \mbox{if } U\in (0,
\infty)  \\
\\
             \end{array}
\right.
\label{betasym}
\end{equation}

\begin{equation}
 \gamma=\sqrt{\frac{1-\beta^2}{2}} ,
\end{equation}

\noindent with

\begin{eqnarray}
A &=& -9g^2+6U(2J-2U+\omega) ~,\\
B &=& 54 g (2JU - g^2 + U\omega)~.
\end{eqnarray}

\noindent The range of variation of the Hamiltonian parameters for
this solution is: $g\in \mathbb{R}-\{0\},\omega \mbox{ and } J \in
\mathbb{R}$.  $U_c$ is the critical value for which the system
undergoes the QPT.

\bigskip

For the non-symmetric phase, $U\in (-\infty, U_c)$, the solutions are

\begin{eqnarray}
 \beta& = &
-\frac{g}{4U}+\frac{(1-i\sqrt{3})A^\prime}{12\sqrt[3]{2} U \left(B^\prime+\sqrt{(B^\prime)^2+4 (A^\prime)^3}\right)^{1/3}}
-\frac{(1+i\sqrt{3})\left(B^\prime+\sqrt{(B^\prime)^2+4(A^\prime)^3}\right)^{1/3}}{24\sqrt[3]{2}U}
~ , \label{betanonsym} \\
 \gamma_R &=&\sqrt{\frac{1-\beta^2}{2}+\frac{\sqrt{-J^2+U^2(\beta^2-1)^2}}{2U}}~ ,\\
 \gamma_L &=& -\frac{J}{2U\gamma_R} ~.
\end{eqnarray}

\noindent with

\begin{eqnarray}
A^\prime &=& 12U(\omega-4U)-9g^2~,\\
B^\prime &=& 54g(2U\omega-g^2)~.
\end{eqnarray}

\noindent For this solution the Hamiltonian parameters are restricted to $g\in
\mathbb{R}-\{0\},\omega \in \mathbb{R} \mbox{ and } J \in \mathbb{R}-\{0\}$.

\bigskip

The critical value of $U$ ($U_c$) can be computed using equations
\eqref{betasym} and \eqref{betanonsym}. The variational parameter
$\beta$ has to be a continuous function of $U$ to ensure the
continuity of the ground state energy. By equating both expressions of
$\beta$ in the symmetric, \eqref{betasym}, and non-symmetric,
\eqref{betanonsym}, phases we can determine numerically the critical
value $U_c$. For instance, for a system defined by $J=1$, $\omega=5$
and $g=5$, the value obtained for $U_{c}$ is $U_{c}=-1.14018$.

\bigskip

A similar analysis can be performed for the limit without coupling to
the molecule obtaining the following expressions for the variational
parameters 
\begin{equation}
\gamma_R=\left\{ \begin{array}{ll}
\sqrt{\frac{1}{2}+\mbox{Sign}(U)\frac{\sqrt{U^2-1}}{2U}}  &  \mbox{if }
U < U_c \\

& \\
\frac{1}{\sqrt{2}} &  \mbox{if } U \geq U_c  \\

\\
             \end{array}
\right.
\label{gammaRnF}
\end{equation}

\begin{equation}
\gamma_L=\left\{ \begin{array}{ll}
\frac{-J}{2U}\frac{1}{\sqrt{\frac{1}{2}+\mbox{Sign}(U)\frac{\sqrt{U^2-1}}{2U}}}
& \mbox{if }
U < U_c \\

& \\
\frac{1}{\sqrt{2}} &  \mbox{if } U \geq U_c  \\

\\
             \end{array}
\right.
\label{gammaLnF}
\end{equation}

In this case, the critical value for the coupling constant is $U_c=-1$.

\begin{figure}[ht]
\captionsetup[subfigure]{labelformat=empty}
\begin{subfigure}[b]{0.45\linewidth}
\centering
\includegraphics[width=\textwidth]{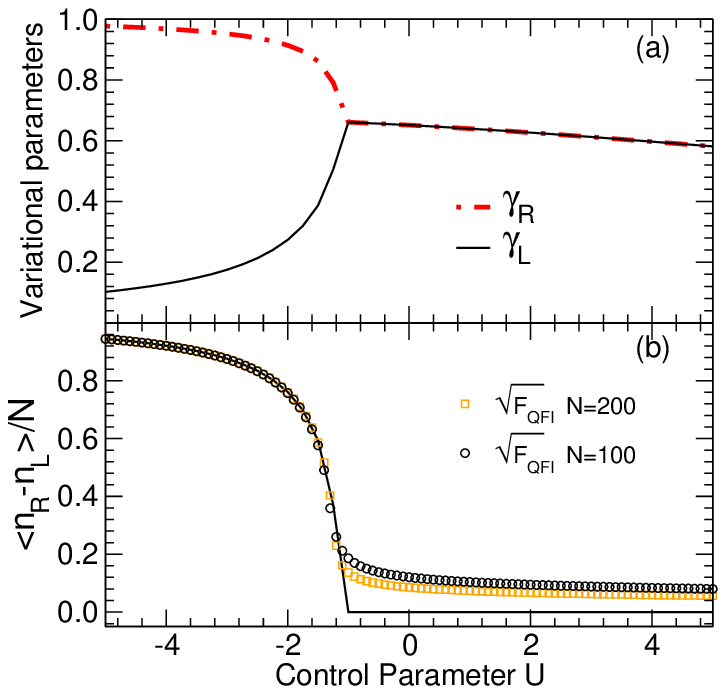}
\label{fig:invalance}
\end{subfigure}
\hspace{0.5cm}
\begin{subfigure}[b]{0.45\linewidth}
\centering
\includegraphics[width=\textwidth]{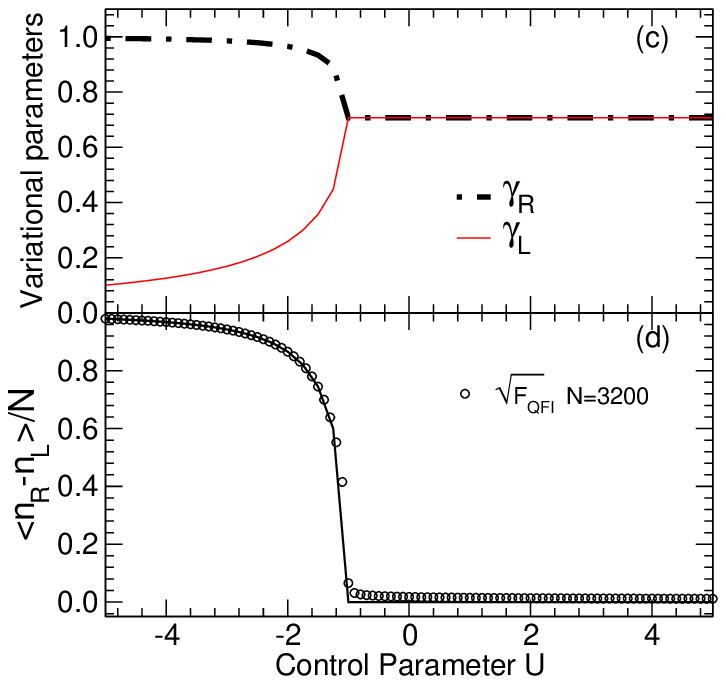}
\label{fig:invalancenF}
\end{subfigure}
\caption{(Color online) Panels (a) and (c) show the variational
  parameters $\gamma_R$ and $\gamma_L$, as a function of the control
  parameter $U$. Panels (b) and (d) show the normalised expectation
  value of the invalance operator $\widehat I$ in solid line, and the
  square root of the Fisher information, obtained by numerical
  diagonalization, in open symbols. Panels (a) and (b) are for the mixed
  atom-molecule system (two numerical calculations for different
  system sizes are shown in (b), (black) circles correspond to
  $N=100$, and (orange) squares, to $N=200$) whereas panels (c) and (d)
  present the system without molecule.}
\label{imbalances}
\end{figure}

Note that Eqs. (\ref{energysurface1})-(\ref{energysurface2}) are
invariant under the transformation $\gamma_L \rightarrow - \gamma_L$,
$\gamma_R \rightarrow - \gamma_R$, which implies that there always
exist two variational solutions. In the symmetric phase, both of them
give rise to $\widehat{I}=0$, corresponding to the same physical
state. On the contrary, in the non-symmetric phase both have non-zero
imbalance, $\widehat{I_1}= - \widehat{I_2}$, implying that the ground
state is degenerate. So, in this case any linear combination of these
two solutions is also a ground-state of the system, and thus it is
possible to build a parity-projected solution, with
$\widehat{I}=0$. As a consequence, the critical coupling $U_c$ separates
two different regions: the symmetric one, in which the populations of
the two wells are always equal, and the non-symmetric one, in which
one can find symmetry-breaking ground states. As the coherent states
used in this section break the symmetry for $\gamma_L \ne \gamma_R$,
they provide an accurate description of the transition in terms of the
imbalance.

\subsection{Numerical results}

In this subsection, we compare the mean field results with exact
diagonalizations for finite systems. In panels (a) and (c)
Fig. \ref{imbalances} we depict the behaviour of the variational
parameters $\gamma_R$ and $\gamma_L$ for both cases, without molecule (right panels) and with molecule (left panels). It is clearly seen
that $\gamma_L$ and $\gamma_R$ are equal to each other in one of the
phases, and different in the other, for both models. In panels (b) and (d)
of the figure we plot the mean-field result for the imbalance,
together with the numerical values for $\sqrt{F_{QFI}}$, obtained for
$N=3200$ particles in the case without molecules, and with $N=100$ and
$N=200$ particles in the case with molecule. The imbalance behaves as
a typical order parameter: it is zero in one of the phases, and
becomes non-zero after crossing the critical point $U_c$, showing a
non-analytic behavior at $U_c$. Therefore, the symmetric phase is
characterized by the same population of atoms in the two wells,
whereas in the non-symmetric phase there is a finite
imbalance. Numerical results show that $\sqrt{F_{QFI}}$ provides also
a correct description of the QPT: it is non-zero in the non-symmetric
phase, and tends to zero in the symmetric one. Note that the Fisher
information measures the fluctuations around the equilibrium state,
and therefore it cannot be strictly zero in finite
systems. However, we can see from the figure how this signature tends
to zero as the size of the system increases.

\begin{figure}[ht]
\captionsetup[subfigure]{labelformat=empty}
\begin{subfigure}[b]{0.45\linewidth}
\centering
\includegraphics[scale=0.33,angle=-90]{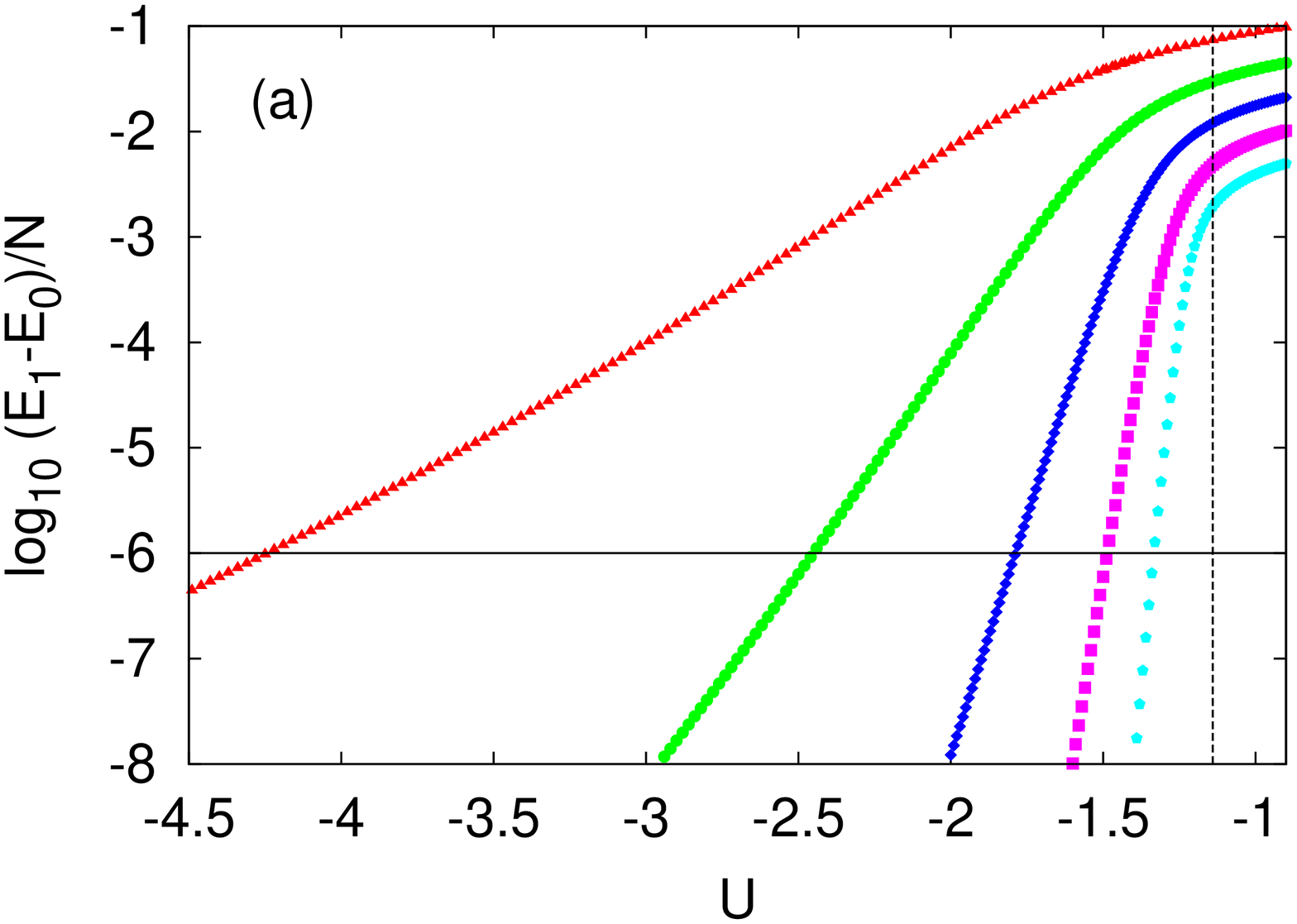}
\label{fig:energy_fes}
\end{subfigure}
\hspace{0.5cm}
\begin{subfigure}[b]{0.45\linewidth}
\centering
\includegraphics[scale=0.33,angle=-90]{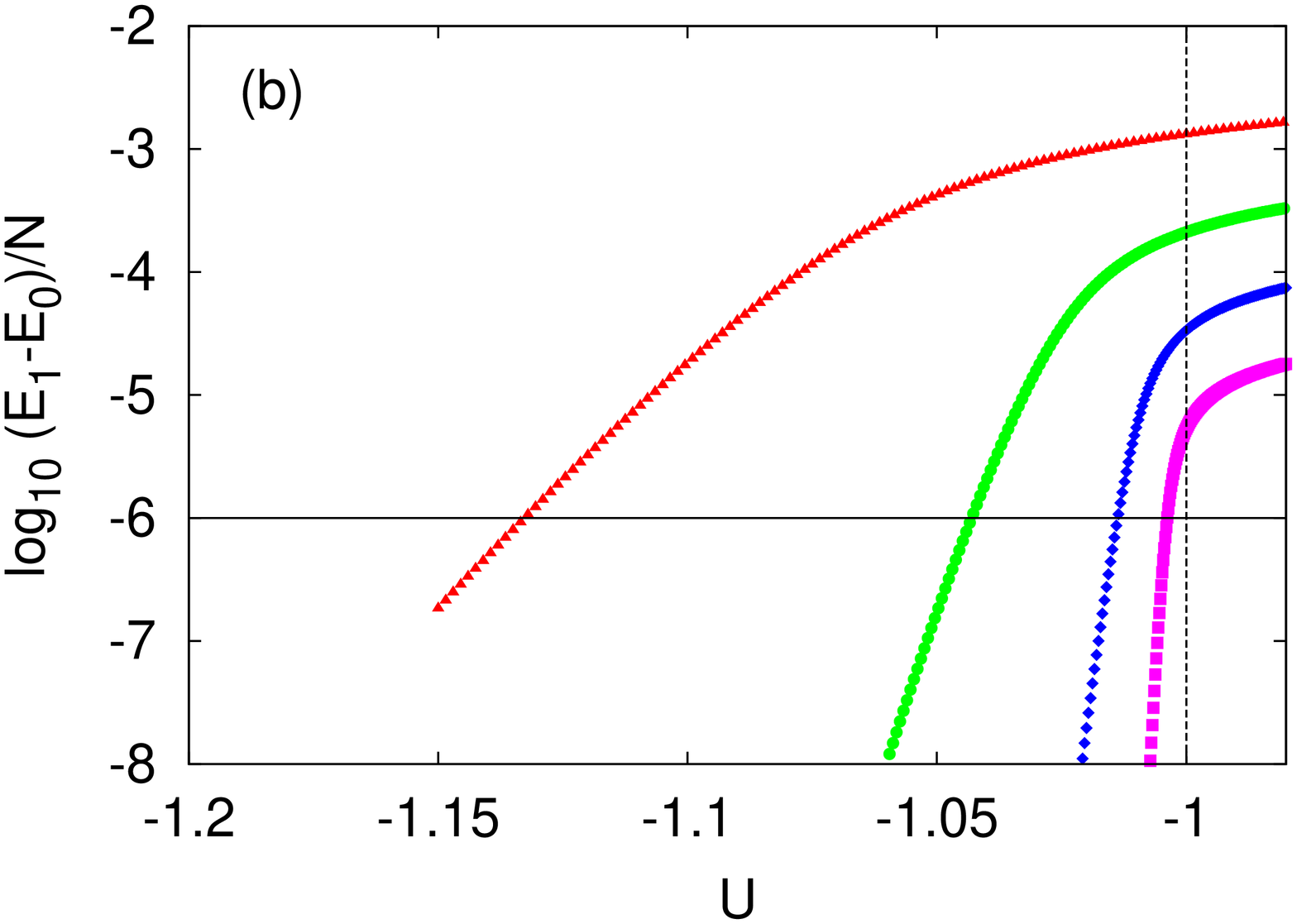}
\label{fig:energy_double}
\end{subfigure}
\caption{(Color online) Gap between ground state and first-excited
  state, for different system sizes: system with molecule, $N= 20$
  (triangles, red online), $40$ (circles, green online), $80$
  (diamonds, blue online), $160$ (squares, magenta online) and $320$
  (hexagons, cyan online), in panel (a); and system without molecule,
  $N= 250$ (triangles, red online), $1000$ (circles, green online),
  $4000$ (diamonds, blue online) and $16000$ (squares, magenta
  online), panel (b). In all the cases, system sizes increase from
  left to right.}
\label{gap}
\end{figure}

In Fig. \ref{gap} we depict the energy (per particle) difference
between the ground state and the first-excited state for both, the
case with (panel (a)) and without (panel (b)) molecule, for
different system sizes. For the former, we have used $N=20$, $40$,
$80$, $160$, and $320$ particles; for the second one $N=250$, $1000$,
$4000$, and $16000$. Dotted vertical lines show the mean field values
for the critical coupling $U_c$. Solid horizontal lines show a
numerical bound to distinguish the degenerate from the non-degenerate
phase. The value of the coupling for which the gap becomes less than
this bound is the finite-size precursor of the transition
$U_c^{(N)}$. Obviously, the precise value of this precursor depends on
the chosen value for the bound, but the important result is the
finite-size scaling and the fact that $U_c^{(N)} \rightarrow U_c$ in
the thermodynamic limit. From our numerical results, it seems clear
that the finite-size precursor of the transition tends to the
mean-field critical behaviour in the thermodynamic limit.

\begin{figure}[ht]
\captionsetup[subfigure]{labelformat=empty}
\begin{subfigure}[b]{0.45\linewidth}
\centering
\includegraphics[scale=0.33,angle=-90]{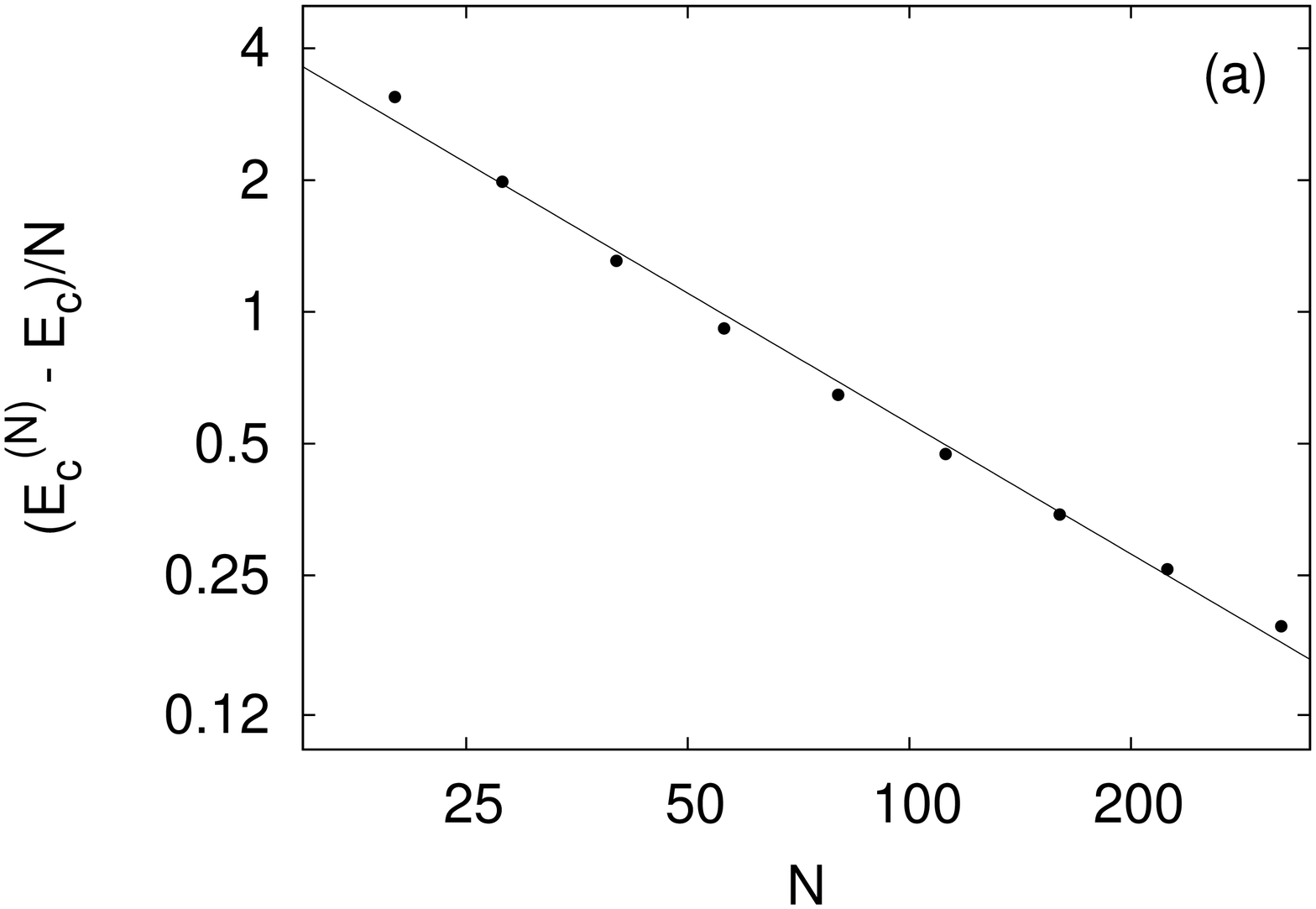}
\label{fig:der2_fes}
\end{subfigure}
\hspace{0.5cm}
\begin{subfigure}[b]{0.45\linewidth}
\centering
\includegraphics[scale=0.33,angle=-90]{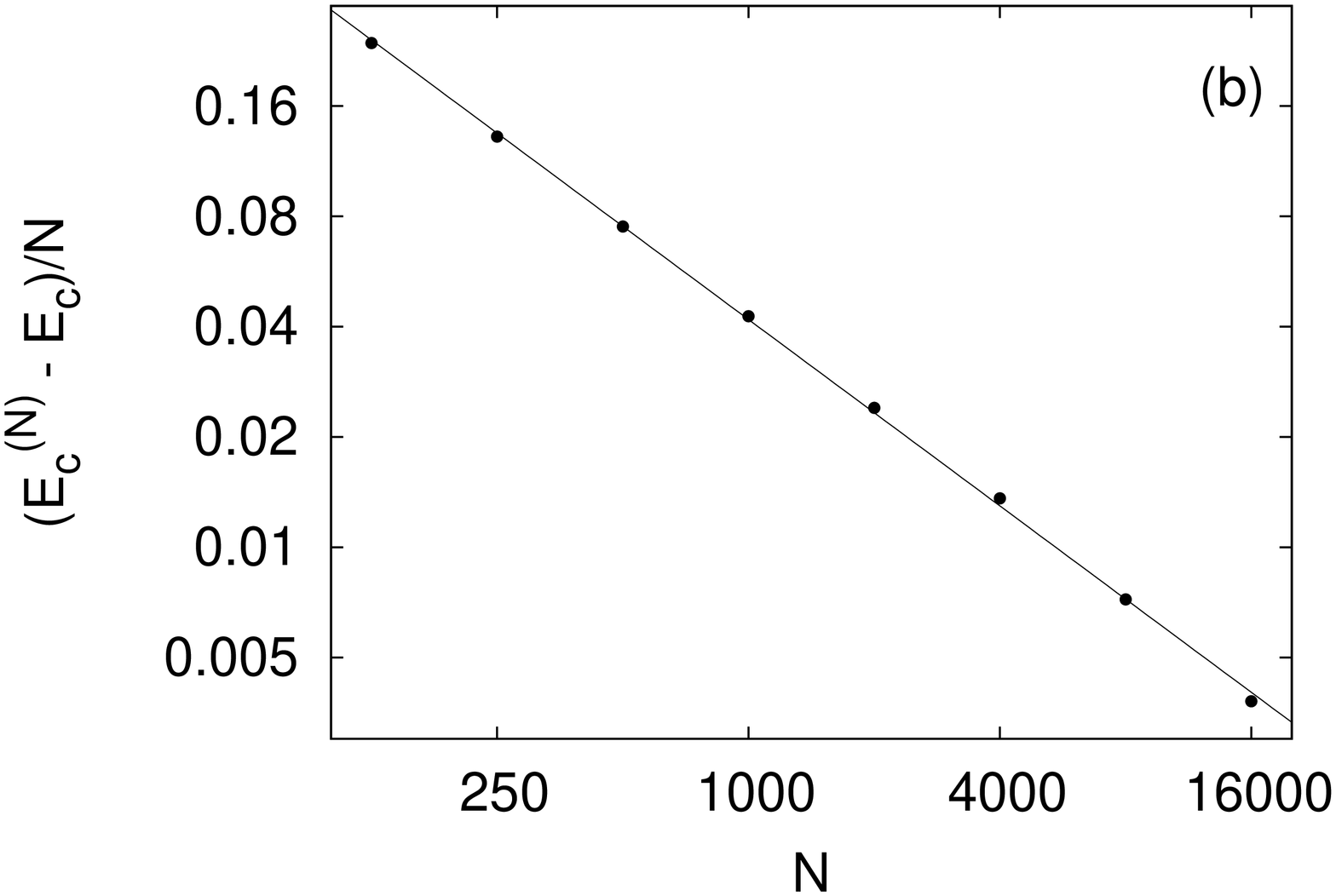}
\label{fig:der2_double}
\end{subfigure}
\caption{Scaling of the finite-size precursor of the critical point.}
\label{scaling_gap}
\end{figure}

In Fig. \ref{scaling_gap} we show how the finite-size precursor
$U_c^{(N)}$ scales with the system size, in a double-logarithmic scale,
both for the case with (panel (a)) and without (panel (b)) molecule. It is
clearly seen that $U_c^{(N)} - U_c \rightarrow 0$, when $N \rightarrow
\infty$. In both cases a scaling power-law $U_c^{(N)} - U_c \propto
N^{-\alpha}$ is found, with $\alpha = 0.84 \pm 0.03$ for the case with
molecule, and $\alpha = 0.99 \pm 0.02$ for the case without molecule.

The results presented in the subsection show that a second-order
quantum phase transition takes place at a critical coupling
$U_c$, for both models. The main difference
between them resides in the finite-size scaling exponent $\alpha$.

\section{Low energy spectrum}

\subsection{Beyond mean-field approximation}

The mean-field results obtained in the preceding section provide the
leading order $N$ description of the models. In order to improve the mean-field
description, we first perform a shift transformation to the $a_L$,
$a_R$ and $b$ bosons, defining the shifted bosons $d_0$, $d_1$ and $f$

\begin{eqnarray}
a^\dagger_R& = &\sqrt{N}\gamma_R + d^\dagger_0, \\
a^\dagger_L& = &\sqrt{N}\gamma_L + d^\dagger_1, \\
b^\dagger_L& = &\sqrt{N/2}\beta + f^\dagger,
\label{shift}
\end{eqnarray}

\noindent where $\gamma_R$, $\gamma_L$ and $\beta$ are taken as real variational
parameters, and $N$ is the total number of particles in the
system. These new bosons,  $d_i$ and $f$, satisfy the usual bosonic commutation relations.

When this shift transformation is introduced into the Hamiltonian, an
expansion of $\widehat {\cal H}$ in powers of $N$ is obtained,
\begin{equation}
\widehat {\cal H} =\widehat H_1 +\widehat H_{1/2} +\widehat H_0 + {\cal O}(1/\sqrt{N})~,
\label{Nexpansion}
\end{equation}
\noindent where $\widehat H_i$ represents the i-esim term of the expansion, which is proportional to the power $N^i$.

The first term, $\widehat H_1$, is the mean-field energy already computed
in the previous section, with $\gamma_R$, $\gamma_L$ and $\beta$
evaluated at the energy minimum. The next order in the Hamiltonian
expansion, $\widehat H_{1/2}$, is

\begin{equation}
\widehat H_{1/2} =
\sqrt{N}\left[\frac{-1}{\sqrt{2}}
\left(\gamma_R^2+\gamma_L^2\right)g+\beta \left(2\lambda-\omega
\right) f - \left( J\gamma_L+(\lambda+\beta
g)\gamma_R -2U\gamma_R^3 \right) (d_0+d_1) + \mbox{h.c.} \right]~.
\label{H_1/2}
\end{equation}

This term cancels at the equilibrium values of the variational
parameters. This stems from the relation $\widehat H_{1/2}
= \sum_{i}\frac{1}{2\sqrt{N}} \frac{d\widehat
H_1}{d\gamma_i}(d_i+d^\dagger_i)+\frac{1}{\sqrt{2N}} \frac{d\widehat
H_1}{d\beta}(f+f^\dagger)$, $i=0,1$. Thus, the
first non-zero correction to the mean-field energy comes from the $\widehat
H_0$ term. This term can be expressed in compact form as

\begin{equation}
\widehat H_0= \frac{1}{2}\delta^\dagger M \delta -\frac{1}{2}\mbox{Tr}Y,
\label{H0}
\end{equation}

\noindent where $\delta$ and $\delta^\dagger$ are column and row vectors
of dimension $6$ respectively, $M$ is a $6\times 6$ matrix, and $Y$ and $Z$ are $3\times
3$ matrices. $M$ has the form

\begin{equation}
M =
 \begin{pmatrix}
  Y & Z \\
  Z & Y
 \end{pmatrix} ,
\end{equation}

\noindent with the $3\times 3$ $Y$ and $Z$ matrices given by

\begin{equation}
Y =
 \begin{pmatrix}
  4U\gamma_R^2-\lambda & -J & -\sqrt{2}\gamma_R g \\
  -J & 4U\gamma_L^2-\lambda & -\sqrt{2}\gamma_L g \\
 -\sqrt{2}\gamma_R g & -\sqrt{2}\gamma_L g & \omega-2\lambda
 \end{pmatrix}  ,
\quad
Z= \begin{pmatrix}
  2U\gamma_R^2-\beta g & 0 & 0 \\
  0 &2U\gamma_L^2-\beta g & 0 \\
 0  & 0 & 0
 \end{pmatrix} ,
\label{matrixes}
\end{equation}

\begin{equation}
 \delta^\dagger =\begin{pmatrix}
          d^\dagger_0 & d^\dagger_1 & f^\dagger & d_0 & d_1 & f
         \end{pmatrix} .
\end{equation}

The quadratic Hamiltonian (\ref{H0}) in the shifted bosons $d$ and $f$
can be diagonalized by means of a canonical transformation
\cite{BlaizotRipka}, which is equivalent to diagonalize the matrix
\begin{equation}
\tilde{M} =
 \begin{pmatrix}
  Y & Z \\
  -Z & -Y
 \end{pmatrix} .
\end{equation}

The matrix $\tilde{M}$ has a Goldstone mode at zero energy due to the
breaking of the $U(1)$ symmetry associated with the condensation
phenomenon. It also has two real and positive eigenvalues describing
the excited states of the system. Once the canonical
transformation is defined through the diagonalization of $\tilde{M}$, $\widehat
H_0$ can be expressed as

\begin{equation}
\widehat H_0= \sum_{i=1}^{3}g^\dagger_{i}g_{i}
\Delta_{i}+\frac{1}{2}\sum_{i=1}^{3}\Delta_{i}-\frac{1}{2}\mbox{Tr}Y,
\end{equation}

\noindent where $g_{i}$ $(g^{\dagger}_{i})$ are the new quasiparticle
operators, and $\Delta_{i}$ ($i=1,2,3$) are the corresponding
eigenvalues of $\tilde{M}$. This problem has no analytic expression
for $\Delta_{i}$ in general. For a given set of Hamiltonian parameters
$g$, $\omega$ and $J$, we diagonalise numerically $\tilde{M}$ to
compute the eigenvalues $\Delta_{i}$. Notice that the term
$\frac{1}{2}\sum_{i=1}^{3}\Delta_{i}-\frac{1}{2}\mbox{Tr}Y$ gives a
correction for the ground state mean field energy.

The limit with no molecule is obtained from the analytic expressions
for the mean field corrections when $g \rightarrow 0$ and
$\omega \rightarrow \infty$.  Notice that the $f$ ($f^{\dagger}$) modes
decouple from the $g_{i}$ $(g^{\dagger}_{i})$ modes and its energy
goes to infinity.  In this limit $\tilde{M}$ becomes a 4x4 matrix and
its eigenvalues provide the two different excited modes of the
model. One of these modes is always zero, corresponding to the
Goldstone mode associated with the breaking of $U(1)$ symmetry.  The
reduction of the $\tilde{M}$ matrix size allows for the
derivation of analytic expressions for these modes since its
characteristic polynomial has degree four and can be solved by
radicals.

\subsection{Comparison with numerical calculations}

\begin{figure}[ht]
\centering
\includegraphics[angle=-90]{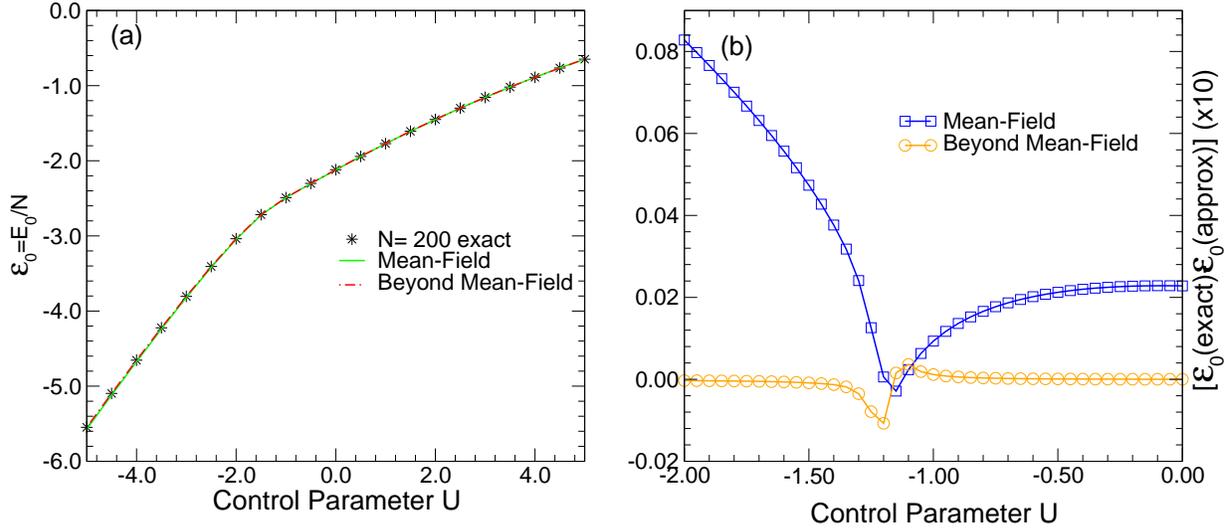}
\caption{\label{figgs}(Color online) In panel (a) the ground
state energy per particle of the model with the diatomic is plotted as a function of the control parameter $U$ for
$N=200$. Panel (b) shows the difference between the exact ground state energy per
particle obtained numerically and the values obtained by using the two
analytic approximations considered in this work in a small region
around the critical point for a system with $N=200$. The rest of the
control parameters in the Bose-Hubbard Hamiltonian are: $J=1$,
$\omega=5$ and $g=5$.}
\end{figure}

In order to check the quality of the mean field approach and the
improvement provided by the corrections to the mean field approximation discussed
in the preceding subsection, we present here a comparison with the
exact diagonalization of (\ref{eq:Hamiltonian}) for finite systems.
The ground state energy per particle versus the control parameter $U$,
for fixed values of the other parameters $J=1$, $\omega=5$, and $g=5$, is
shown in panel (a) of Fig.\ \ref{figgs}. The stars represent the exact
numerical calculation for a system with $N=200$ bosons. The thin full
line is the mean-field result, while the dot-dashed line corresponds
to the beyond mean-field approximation. For this system size the
differences between the exact results and the two approximations are
imperceptible at the panel (a) scale. In order to see the numerical
differences, we present in panel (b) the differences between the
mean-field and beyond mean-field calculations with the exact numerical
results in a larger scale. The $U$ values in (b) are restricted to the
region around the critical point at $U_c \sim -1.14$. As apparent in
Fig.\ \ref{figgs} (b), the computed ground
state energies from beyond mean field clearly improve the mean field
results. The beyond mean field calculations match the exact results in
most of the range of variation of the control parameter; the
differences with the exact numerical results are concentrated in a
small region close to the critical point. This validates the beyond
mean field formalism as a valuable tool for taking into account the
finite-size effects in these models.

\begin{figure}[ht]
\begin{subfigure}[b]{0.45\linewidth}
\centering
\includegraphics[angle=-90, width=1\textwidth]{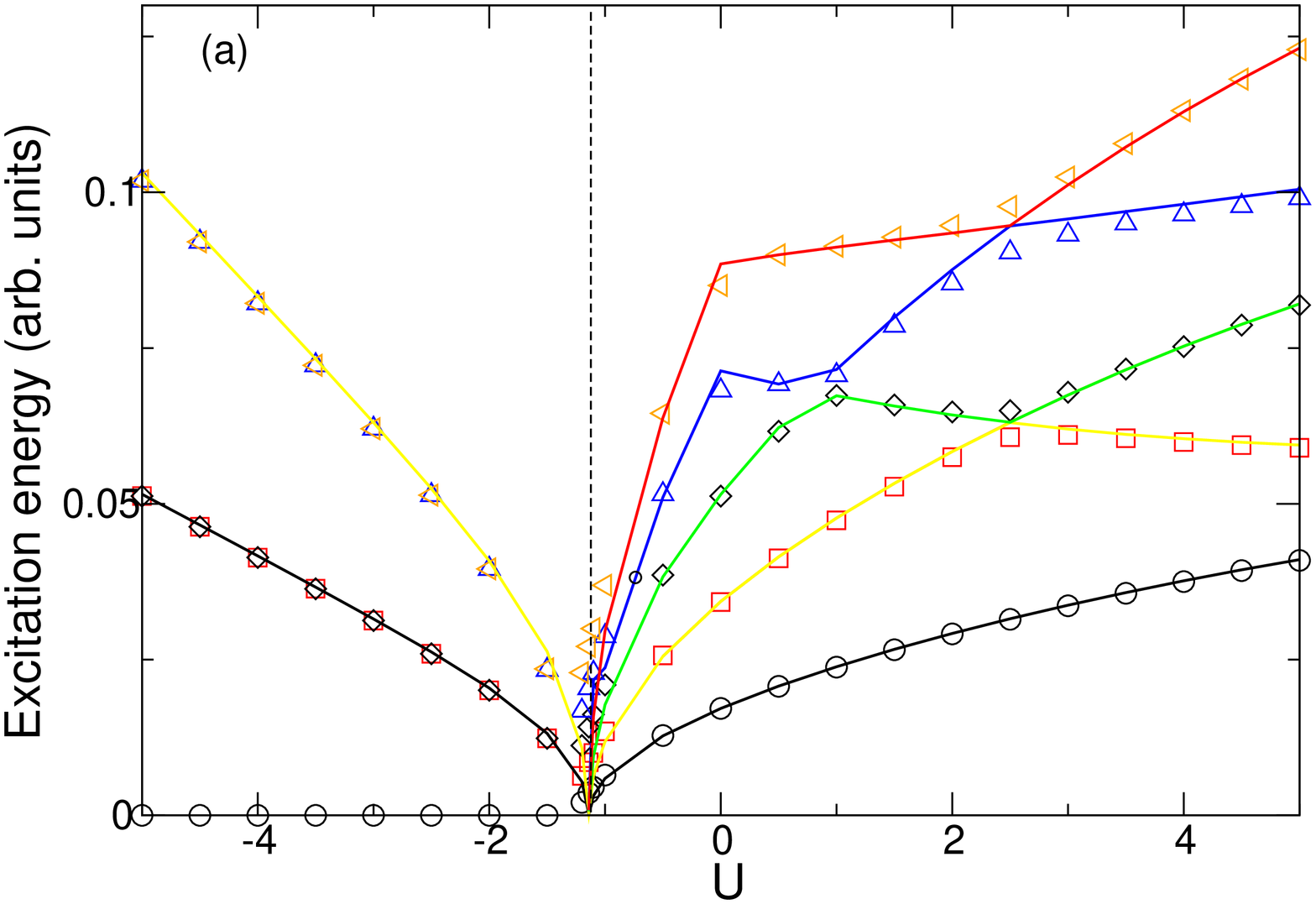}
\end{subfigure}
\hspace{0.5cm}
\begin{subfigure}[b]{0.45\linewidth}
\centering
\includegraphics[angle=-90, width=1\textwidth]{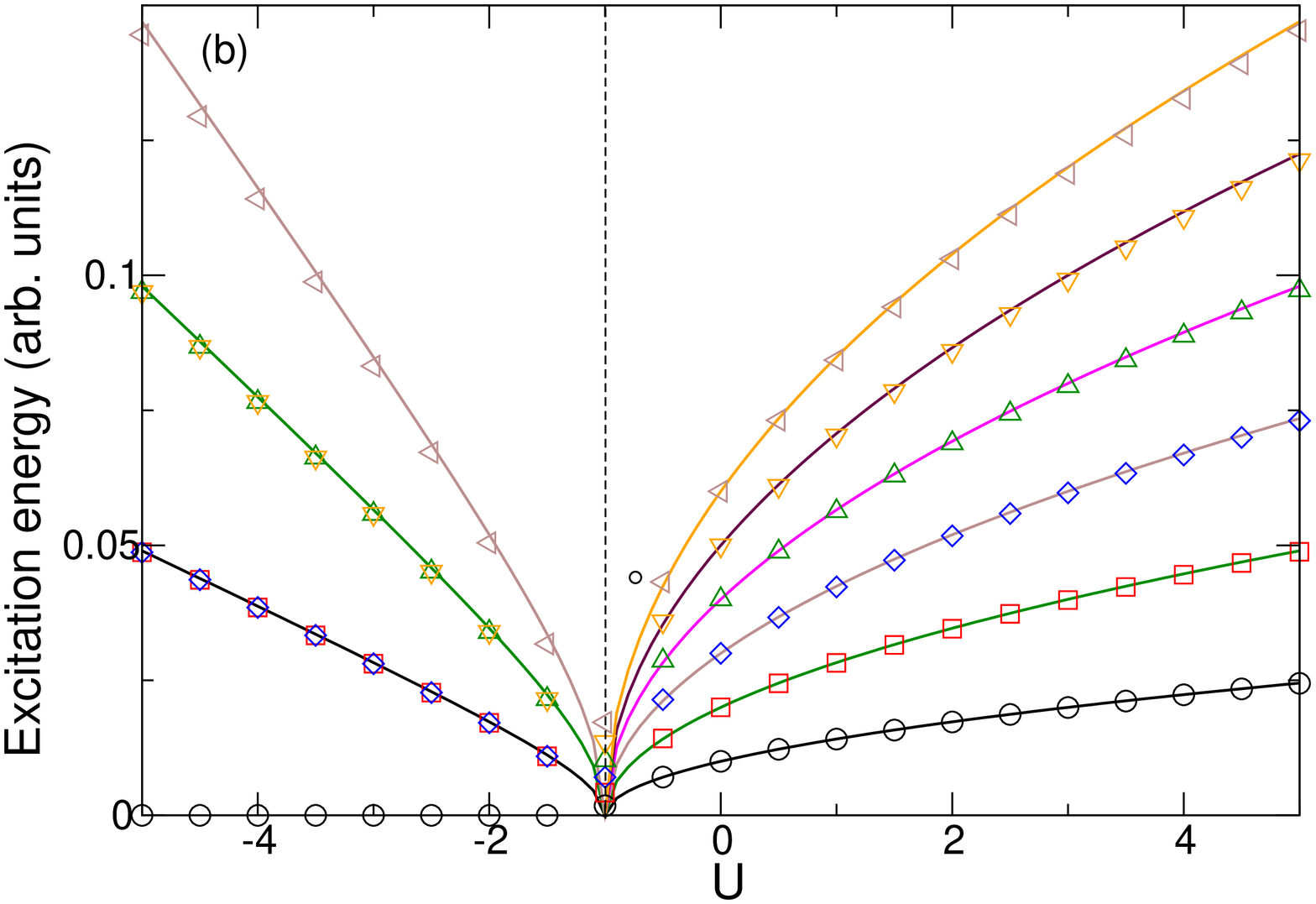}
\end{subfigure}
\caption{(Color online) First few excited state energy gaps as a function
of the control parameter $U$. Symbols are the exact results for $N=200$,
while lines provide the beyond mean field approximation. The rest of
the parameters are fixed to $J=1$, $\omega=5$ and $g=5$ in panel (a), and $J=1$ in panel (b). The vertical dot line marks the
separation between the two phases.}
\label{figgap}
\end{figure}

Fig. \ref{figgap} shows the lowest excited energies obtained by means
of the beyond mean-field treatment (full lines), compared with the
exact numerical results (symbols), for both the case with (panel (a))
and without (panel (b)) the diatomic molecule. In both cases, the
beyond mean-field approach gives a very accurate description of the
low-energy spectra, including the important fact that the energy
levels are degenerate in pairs in one of the phases, while this
degeneracy is broken in the other. The main difference between the two
systems lay in the qualitative behavior of the levels. In the case
without the molecule, the spectrum is similar in both phases, though it
seems more compressed in the symmetric one. On the contrary, both
phases show clear qualitative differences in the case with the
molecule. The spectrum of the symmetry-broken phase is very smooth,
all the levels changing mildly with energy. The symmetric phase is
characterized by an erratic behavior of the levels, with crossings
between levels with different parities, and multiple avoided
crossings. This constitutes a signature of a highly complex behavior
related to the onset of chaos (see Sec. V for more details).

\begin{figure}[ht]
\captionsetup[subfigure]{labelformat=empty}
\begin{subfigure}[b]{0.45\linewidth}
\centering
\includegraphics[angle=-90, width=1\textwidth]{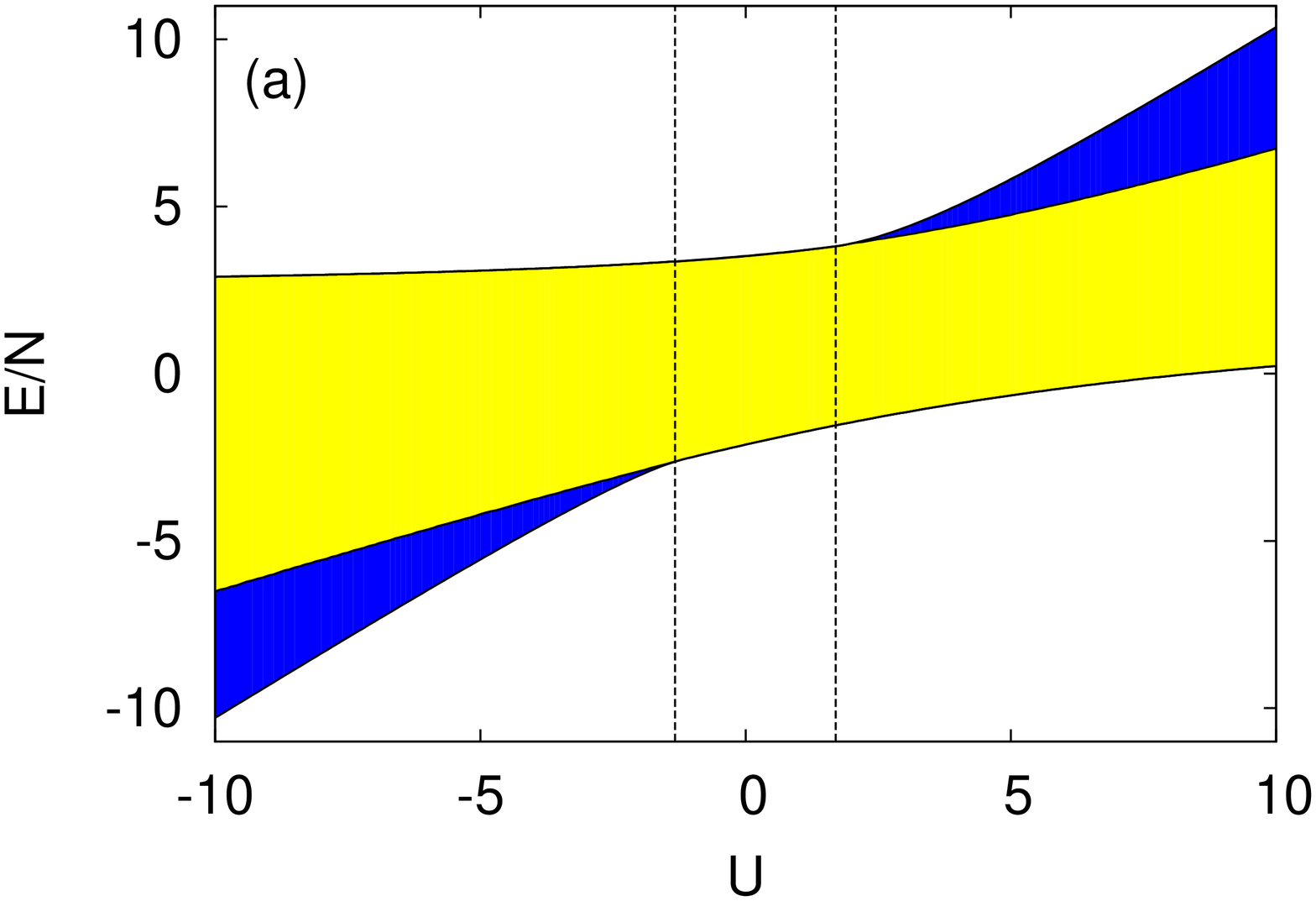}
\label{diagram}
\end{subfigure}
\hspace{0.5cm}
\begin{subfigure}[b]{0.45\linewidth}
\centering
\includegraphics[angle=-90, width=1\textwidth]{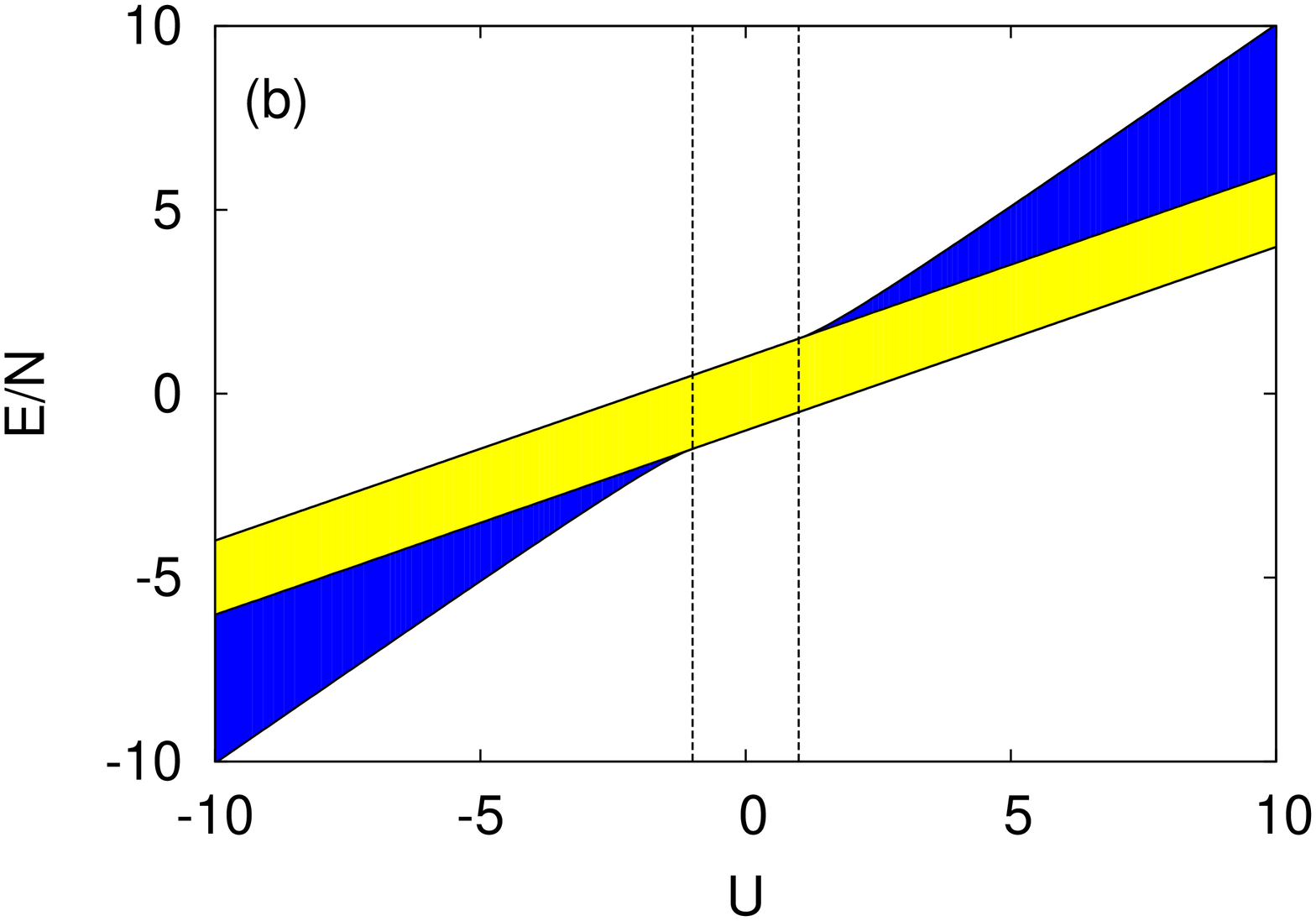}
\label{diagramnF}
\end{subfigure}
\caption{(Color online) Complete phase diagram. Panel (a) depicts the
  case with molecule, obtained with $N=320$ particles, and $J=1$,
  $\omega=5$ and $g=5$. Panel (b) depicts the case without molecule,
  with $N=1000$ particles and $J=1$. In both cases, dark filled region
  (blue online) indicates the symmetry-breaking phase, in which the
  eigenstates are doubly-degenerate, and light filled region (yellow
  online) the normal phase, in which there are no degeneracies and the
  parity is a good quantum number for all the eigenstates. Vertical
  dotted lines enclose the region in which there is no ESQPT.}
\label{ESQPT}
\end{figure}

\section{Excited-state quantum phase transitions}

In some collective many-body quantum systems, QPTs are accompanied by
ESQPTs, giving rise to a critical energy $E_c$ for certain values of
the system control parameter
\cite{Relano:08,Pedro:11b,Puebla:13,Arias:03,Cejnar:10,Ribeiro:08,Perez-Bernal:10,Pedro:11,Brandes:13}. Traditionally,
  ESQPTs have been linked to singularities in the density of states or
  in one of its derivatives, depending on the number of system's
  degrees of freedom in the semiclassical limit \cite{Pavel:14}. It
  has been shown that for a system with just one effective degree of
  freedom, as the Tavis-Cummings model, a $\lambda$-singularity in the
  density of states characterizes the ESQPT. For the Dicke
  model, which has two effective degrees of freedom, the singularity
  making the ESQPT appears in the first derivative of the density of
  states \cite{Brandes:13}. This fact entails an increasing difficulty
  for detecting such a critical behavior in complex systems, with many
  effective degrees of freedom.

  In this work, we rely on a different way to identify ESQPTs. In
  general, the critical energy separates two different regions of the
  spectrum which share some of the properties of the ground state at
  one and at either side of the QPT. For instance, in Dicke and Lipkin
  models the critical energy separates one region in which all
  eigenstates are doubly-degenerate from another in which there are no
  degeneracies and all the levels have a well defined parity
  \cite{Puebla:13}. So, in order to characterize the ESQPTs in our
  system, we study the distance between consecutive energy levels with
  opposite parity. We follow the same criterion as in
  Fig. \ref{gap}. Figure \ref{ESQPT} shows the energy per particle
  versus the strength of the $U$ parameter for the two systems. Panel
  (a) corresponds to the case with a diatomic molecule and a total of
  $N=320$ particles, with $J=1$, $\omega=5$ and $g=5$; while panel (b)
  corresponds to the case without molecule, with $N=1000$ particles
  and $J=1$. The lowest line is the ground-state energy, while the
  uppermost line gives the limit of the calculated spectra for the
  system size used (highest calculated excited state). There are many
  excited states in between both lines. In both panels two vertical
  dotted lines are plotted: the left line gives the ground state QPT,
  $U_c$, while the right one marks an equivalent singularity for the
  highest excited state, $\widetilde{U}_c$.  The energy region with no
  degeneracies is called normal (light filled region, yellow online),
  while the region with double degeneracies is a reflection of the
  appearance of a broken-symmetry (dark filled region, blue
  online). Our calculations clearly identify an ESQPT, that is marked
  by the line in between light and dark regions. It separates normal
  (non-degenerate) from broken-symmetry (degenerate) phases.  Figure
  \ref{ESQPT} shows that both systems behave in the same qualitative
  way. For $U<U_c$, the lower part of the spectrum is degenerate,
  whereas the upper part is not degenerate. For large values of the
  control parameter, $U>\widetilde{U}_c$, the spectrum behaves in the
  opposite way: the upper part is degenerate, whereas the lower part
  is not. In between, $U_c \leq U \leq \widetilde{U}_c$, all the
  energy levels are not degenerate. This entails that, for attractive
  interactions $U<0$ there is a QPT at a critical coupling $U_c$, and
  a critical energy $E_c$ for $U<U_c$. On the contrary, if the
  interaction is repulsive $U>0$, an analogue of a QPT takes place on
  the highest-excited level at a critical coupling $\widetilde{U}_c$,
  and also a critical energy appears for $U>\widetilde{U}_c$. No ESQPT
  is observed in the region $U_c \leq U \leq \widetilde{U}_c$.

The critical line can be easily estimated from the
numerics in the case without the molecule and gives
\begin{equation}
E_c =
\begin{cases}
\frac{U}{2} - 1, & \, U < U_c=-1; \\
\frac{U}{2} + 1, & \, U > \widetilde{U}_c=1. \\
\end{cases}
\label{eq:esqpt_sin}
\end{equation}
This result entails that the spectrum is turned upside down when
changing the interaction from attractive to repulsive, being the
critical points symmetric with respect to the free case,
$U=0$. The same analysis is not that clean for the system
with a molecule. The precise values for $\widetilde{U}_c$ and $E_c
(U)$ depend on the molecular parameters $\omega$ and $g$. For $\omega \ne 0$ and $g \ne 0$ the
symmetry around $U=0$ is broken. Furthermore, due to the smaller
values of the total number of particles $N$ accessible to the our
current computational power, it has not been possible to
obtain a precise estimate of the behavior in the thermodynamic
limit. For $g=\omega=5$ a numerical fit to the case with $N=320$
particles gives rise to
\begin{equation}
E_c =
\begin{cases}
0.46 U - 1.89, & \, U < U_c \sim -1.3 \\
0.42 U + 2.55, & \, U > \widetilde{U}_c \sim 1.7 \\
\end{cases}
\label{eq:esqpt_con}
\end{equation}
where both critical values, $U_c$ and $\widetilde{U}_c$, correspond to
the finite-size precursors, which are still far from the true critical
points in the thermodynamic limit.

\bigskip

It is possible to  study ESQPT for fixed values of the system parameters
by looking at the density of states $\rho(E)$ as a function of the
excitation energy. A singular behavior in $\rho(E)$ signals an ESQPT
at the critical energy $E_c$. It
has been recently shown that $\rho(E)$ presents a logarithmic
singularity at $E=E_c$ for models with just one semiclassical degree
of freedom. On the contrary, the density of states for models with
more than one semiclassical degree of freedom is not
singular. In these cases, the logarithmic singularity occurs in its
derivative $\rho'(E)$ \cite{Pavel:14}. This feature has
been tested in the integrable Tavis-Cummings model with one
semiclassical degree of freedom, and 
in the non-integrable Dicke model with two semiclassical degrees of freedom
\cite{Brandes:13}. Following a similar procedure, we fix the value of the
interaction $U=9$ and calculate the density of states as a function of the
excitation energy for the two systems.  From the
results of Eqs. (\ref{eq:esqpt_sin}) and (\ref{eq:esqpt_con}) we
estimate the critical energy to be around $E \sim 6.3$ in the case
with molecule, and $E \sim 5.5$ in the case without molecule.

\begin{figure}[ht]
\captionsetup[subfigure]{labelformat=empty}
\begin{subfigure}[b]{0.45\linewidth}
\centering
\includegraphics[scale=0.33,angle=-90]{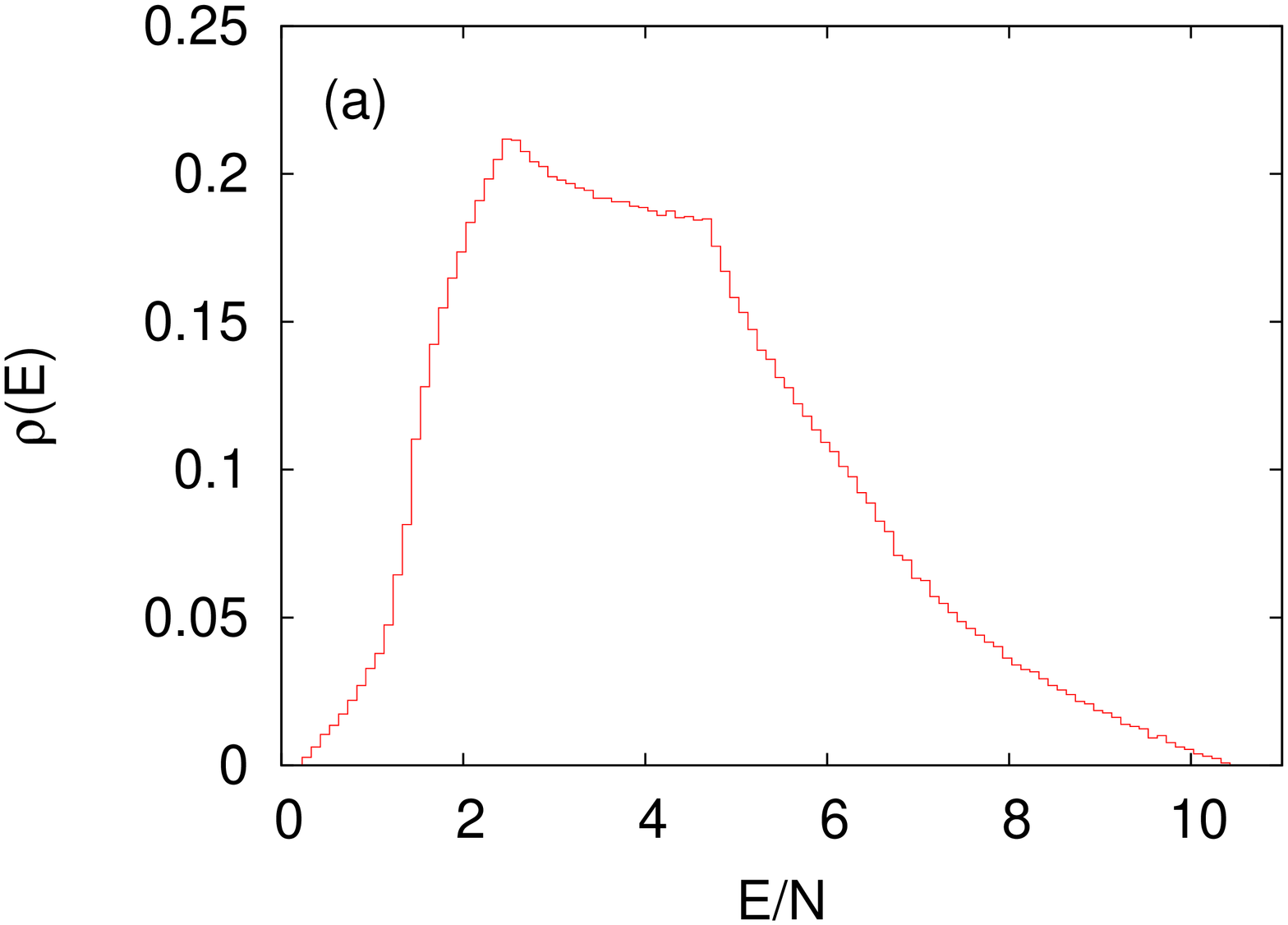}
\label{fig:density_fes}
\end{subfigure}
\hspace{0.5cm}
\begin{subfigure}[b]{0.45\linewidth}
\centering
\includegraphics[scale=0.33,angle=-90]{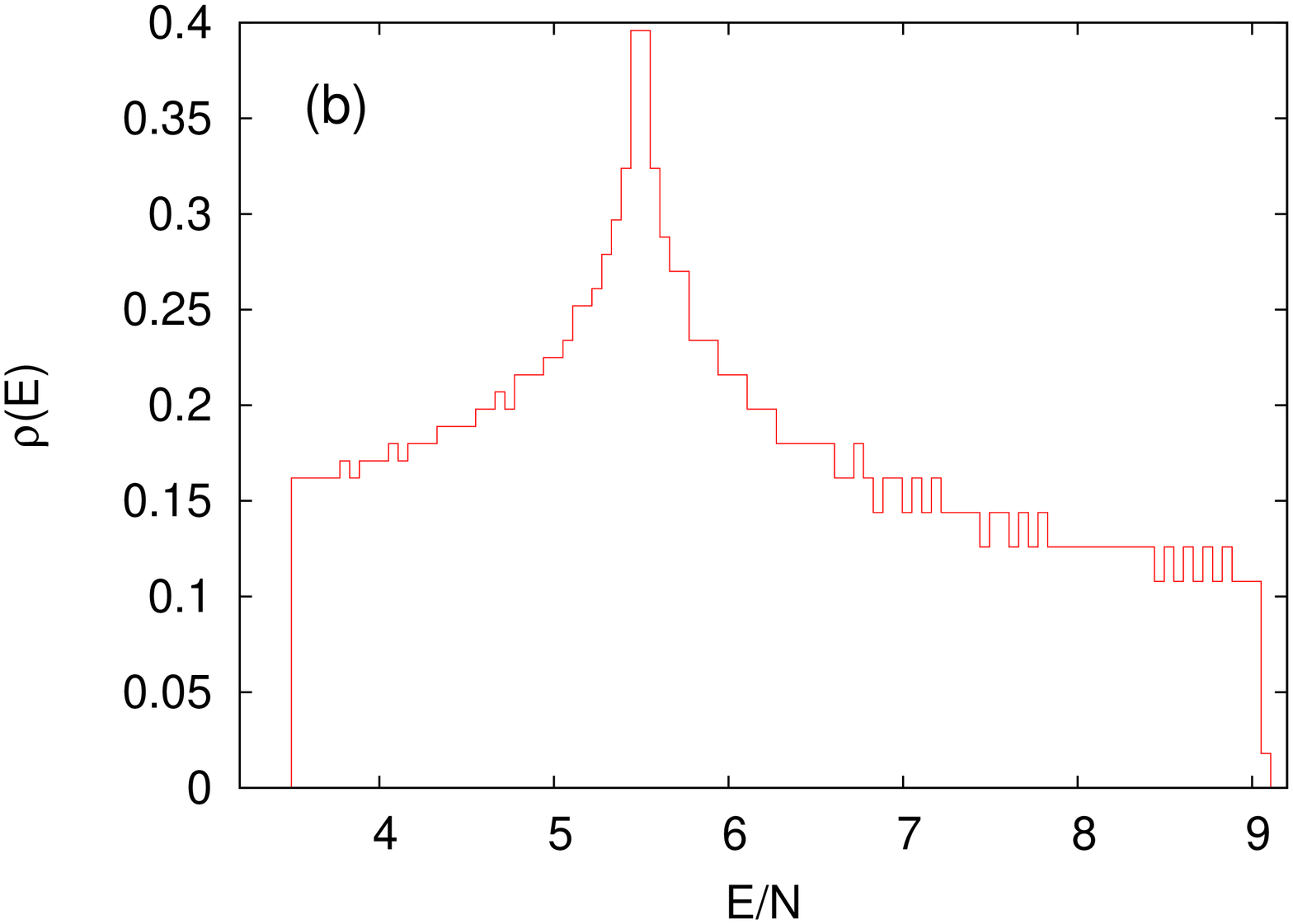}
\label{fig:density_nofes}
\end{subfigure}
\caption{Density of states $\rho(E)$: Panel (a) for $N=320$, $U=9$
  and a molecule at resonance. Panel (b) for $N=2000$ without
  molecule.} 
\label{densidades}
\end{figure}

Figure \ref{densidades} shows the density of states $\rho(E)$ versus
the excitation energy. Panel (a) corresponds to the case of a molecule
at resonance and $N=320$ particles. It shows no neat trace of a
singular behavior around the estimated critical energy $E_c \sim
6.3$. This is however compatible with a logarithmic singularity in the
derivative of the density of states, which is very difficult to see in
systems of this size. Panel (b) corresponds to $N=2000$ without
molecule. It clearly shows a precursor of the logarithmic singularity
at $E_{c}=5.5$. The results obtained with these two systems are
therefore, compatible with those derived in Ref. \cite{Brandes:13} for
the Tavis-Cummings and Dicke models.

\bigskip

\begin{figure}[ht]
\captionsetup[subfigure]{labelformat=empty}
\begin{subfigure}[b]{0.45\linewidth}
\includegraphics[scale=0.33,angle=-90]{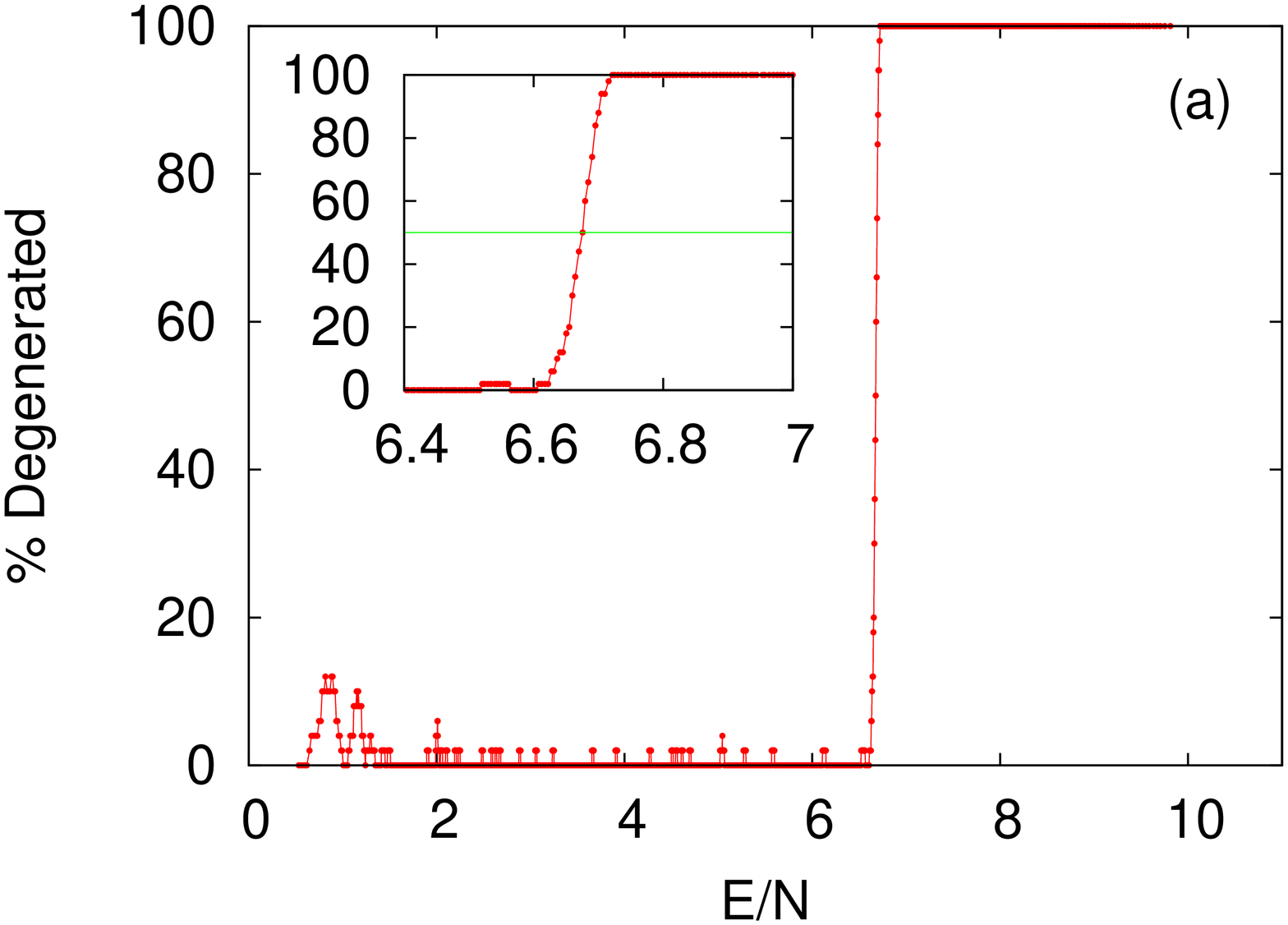}
\label{fig:degenerate_fes}
\end{subfigure}
\hspace{0.5cm}
\begin{subfigure}[b]{0.45\linewidth}
\includegraphics[scale=0.33,angle=-90]{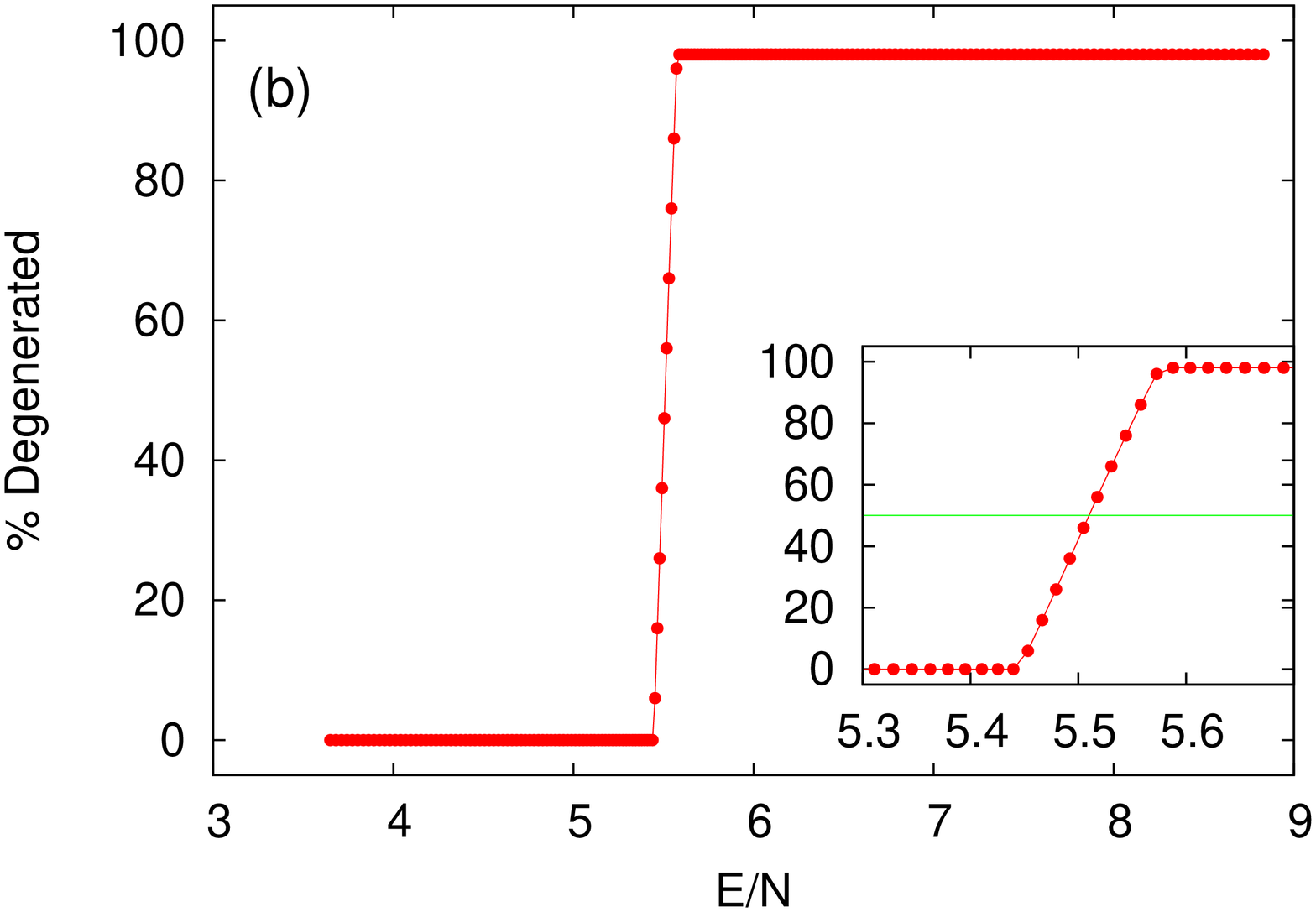}
\label{fig:degenerate_nofes}
\end{subfigure}
\caption{Ratio of degenerate energy levels as a
function of the energy: panel (a) for $N=320$, $U=9$ and a molecule
at resonance. Panel (b) for $N=2000$ without molecule.} 
\label{degeneracion}
\end{figure}

In order to get a deeper insight into the properties of the ESQPTs, we plot in
Fig. \ref{degeneracion} the ratio of degenerate energy levels as a
function of the scaled energy $E/N$. The calculation has been done in
the following way: a) we choose a set of 100 energy levels,
starting from the ground state; b) we count the number of pairs
whose relative energy difference is less than $10^{-6}$; c) we
normalize this value to $100$; d) we associate this value to an
energy equal to the mean energy of the interval; e) we repeat
the calculation starting from the 11$^{th}$ energy level; f) we 
proceed in the same way up to covering the whole spectrum. In the same
figure we plot  
the results for the system with (panel (a)) and without
the molecule (panel (b)). The case without molecule shows 
an abrupt transition from non-degenerate to degenerate energy
levels at the expected value for the critical energy $E_c \sim
5.5$. The case with molecule is not as sharp, though the critical
behavior is still clearly observed. For this case, a small number of
degeneracies appear distributed randomly in the lower
part of the spectrum. We attribute this behavior to the spectral
fluctuations of the energy levels, and relate it to the 
crossings and avoided crossings shown in Fig. \ref{figgap} in the
symmetric phase. In spite of this behavior at low energies, we observe
an abrupt change from non-degenerate to degenerate in the spectrum at
an energy $E_c \sim 6.6$. The $5\%$ discrepancy with our estimate is
attributed to finite-size effects.

\bigskip

\begin{figure}[ht]
\captionsetup[subfigure]{labelformat=empty}
\begin{subfigure}[b]{0.45\linewidth}
\includegraphics[scale=0.33,angle=-90]{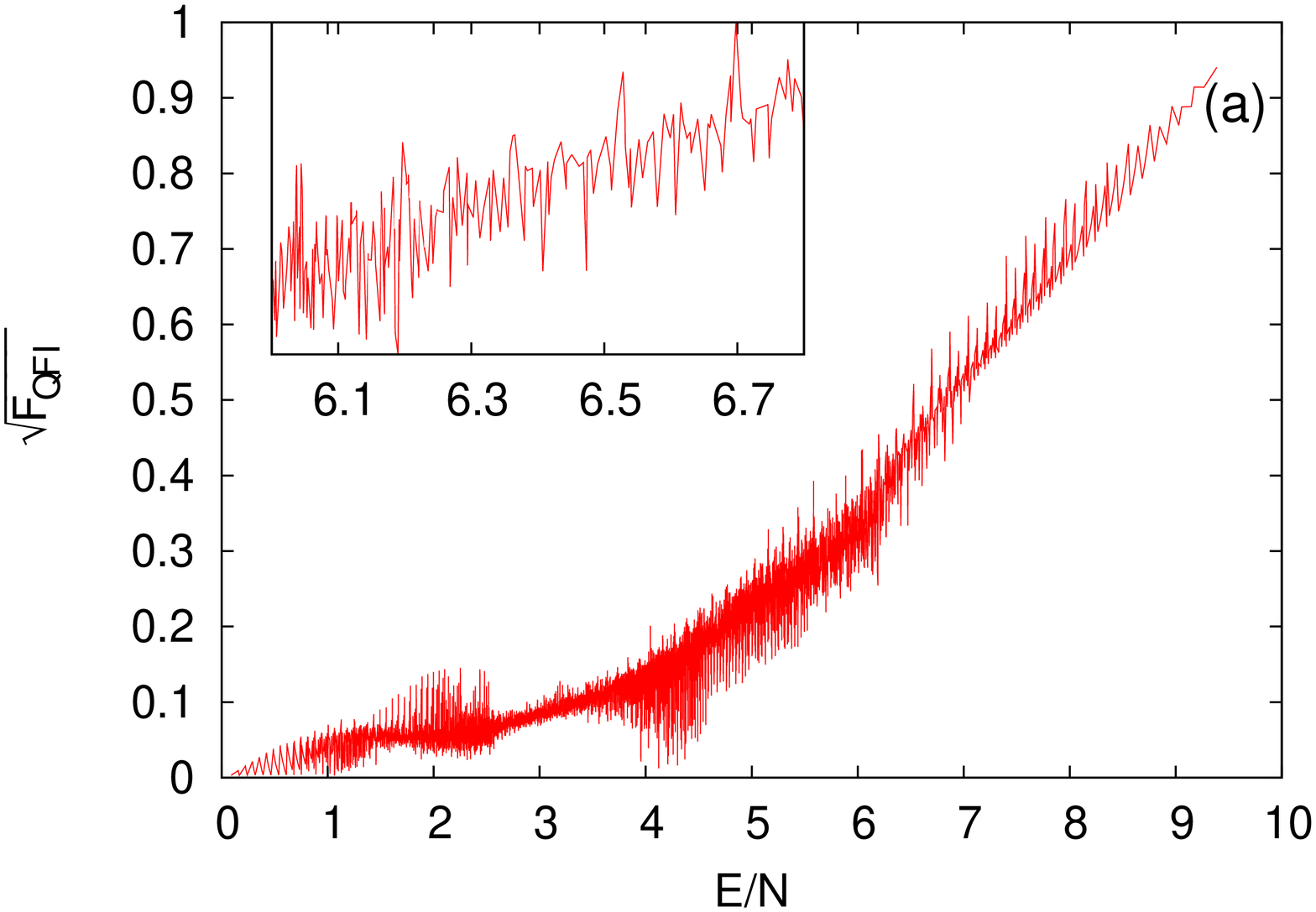}
\label{fig:fisher_excited_fes}
\end{subfigure}
\hspace{0.5cm}
\begin{subfigure}[b]{0.45\linewidth}
\includegraphics[scale=0.33,angle=-90]{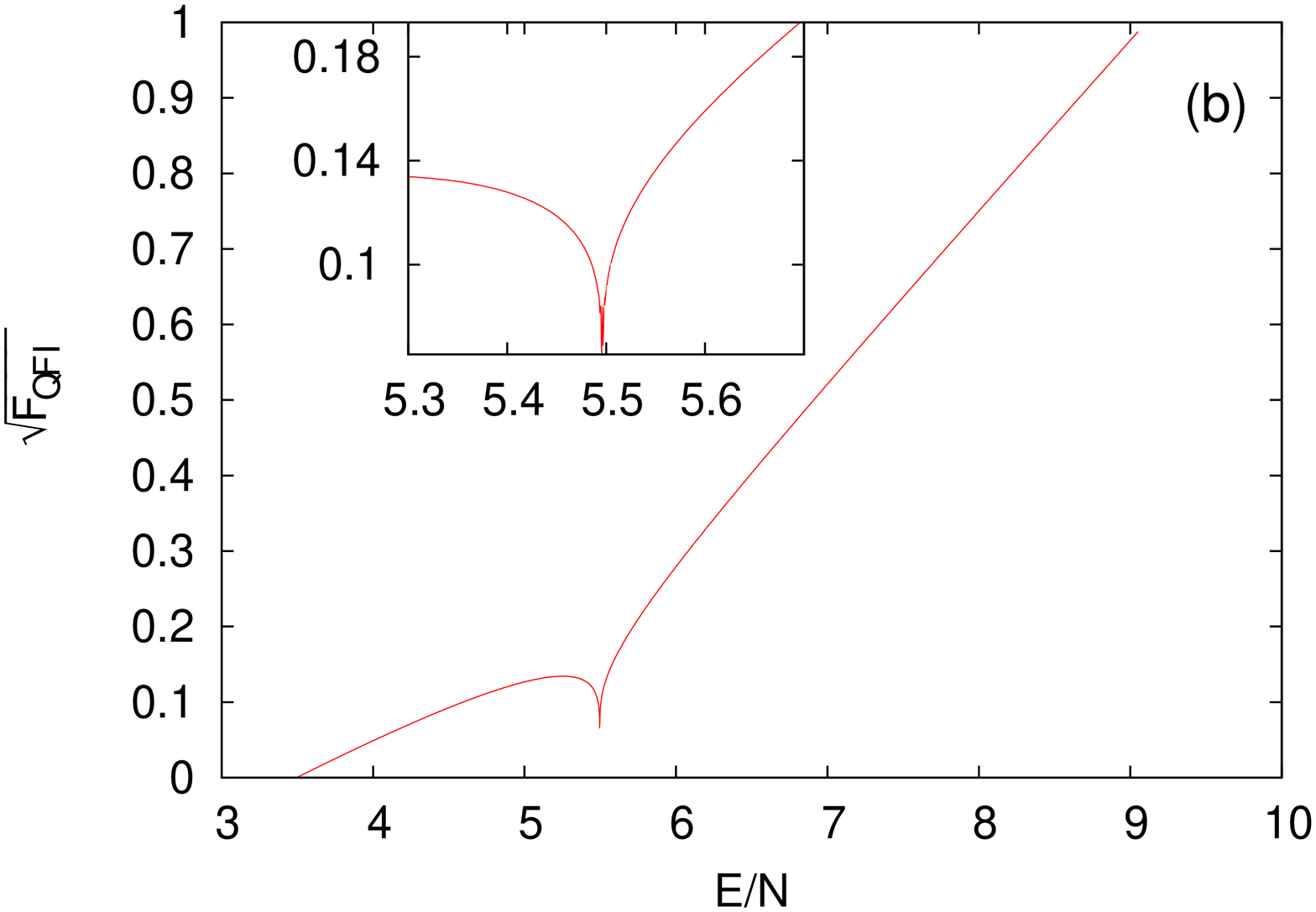}
\label{fig:fisher_excited_nofes}
\end{subfigure}
\caption{Fisher information as a function of the
energy:  Panel (a) for $N=320$, $U=9$ and a molecule at resonance. Panel (b) for $N=2000$ without molecule.}
\label{esqpt_fisher}
\end{figure}

Figure \ref{esqpt_fisher} shows the behavior of the order parameter
$\sqrt{F_{QFI}}$ for both systems, with molecule (panel (a)) and without
molecule (panel (b)). It has been obtained by calculating the
  square root of the Fisher information for all the eigenstates of the
  systems. The results are similar to those obtained for the density
of states. For the case without the molecule the quantum Fisher
information displays a sharp singularity at the estimated critical
energy $E_c \sim 5.5$. On the contrary, there is no such a signature
in the case with molecule. These results are compatible with the fact
that ESQPTs are softer in systems with more than one semiclassical
degree of freedom, as it is pointed out in \cite{Pedro:11,
  Pavel:14}. In the case with the molecule, large fluctuations between
different eigenstates blur the expected singular behavior of the order
parameter at the estimated critical energy. On the other hand, in the
limit without molecule, which has just one semiclassical degree of
freedom, a cusp singularity at the critical energy is clearly seen.

\bigskip

\begin{figure}
\includegraphics[scale=0.33,angle=-90]{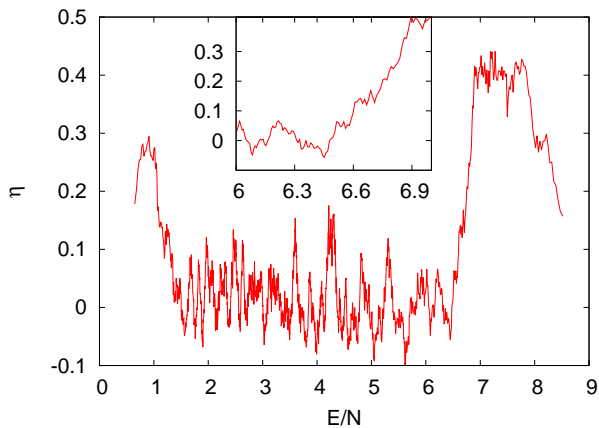}
\caption{Spectral statistics for $U=9$ and $N=360$, even-parity
  states, as a function of the scaled energy $E/N$.}
\label{fig:chaos_fes}
\end{figure}

Finally, it is interesting to analyze the degree of chaos in the
spectra as a function of the energy. In Ref. \cite{Pedro:11b} it was
conjectured a connection between the ESQPT and the onset of chaos in
the Dicke model. Here we analyze the nearest-neighbor spacing
distribution as the simplest measure of quantum chaos. While
integrable systems follow a Poisson distribution $P_P (s) = 
\exp(-s)$, non-integrable and fully chaotic systems are well described by the
Wigner surmise $P_W (s) = \frac{\pi}{2} s \exp \left( - \frac{\pi}{4}
  s^2 \right)$. Given a set of spacings $s_i = \left(E_{i+1} - E_i
\right)/\left< s \right>$, a natural way to measure the degree of
chaos is by means of the ratio
\begin{equation}
\eta = \frac{\text{var}_s - \text{var}_W}{\text{var}_P - \text{var}_W},
\end{equation}
where ``$\text{var}$'' means the variance of the distribution. The
parameter $\eta$ varies from $\eta=0$ for fully chaotic systems, to
$\eta=1$, for integrable systems.

\medskip

In Fig. \ref{fig:chaos_fes} we show the parameter $\eta$ as a function of the
scaled energy $E/N$. The calculation has been done as
follows: a) a set of 250 levels, starting from the ground state has
been chosen, b) the set has been unfolded, discarding a $20\%$ of
higher and lower levels, c) the parameter $\eta$ has been calculated,
d) the value of $\eta$ has been linked to the mean energy value of the
set, e) the procedure has been repeated increasing the energy. It is
important to note that all the eigenstates have a definite value of parity. 
As it can be seen in Fig. \ref{fig:chaos_fes}, a transition
from a fully chaotic to an intermediate regime takes place at $E \sim
E_c$. Below this critical point, the system is more or less chaotic,
with the exception of the region very close to the ground state, which
is closer to integrability. This is fully compatible with the complex
behavior shown in Fig. \ref{figgap} for the lowest excited states in
the symmetric phase. On the contrary, above the critical energy the
system is far from chaos. The interplay between chaos and regularity
as a function of energy is qualitatively similar to the Dicke model
\cite{Pedro:11b,Hirsch:14}. For  
Dicke and the present model, chaos characterizes the normal phase, in which
there are no degeneracies, whereas the symmetry-breaking phase is
closer to integrability. Despite this fact, there seems to be no clear
connection between the ESQPT and the onset of chaos
\cite{Hirsch:14}. Dynamics of systems with two semiclassical degrees
of freedom suffering ESQPTs seems to be more chaotic in the phase in
which there are no degeneracies in the spectrum.


\section{Conclusions}

By means of analytic and numerical results, we have studied the QPT
signaled by the population imbalance in two kinds of double-well BECs:
one comprising bosonic atoms tunneling between the two wells, and the
second including the interaction of the atoms with a diatomic
molecule. Our main results are summarized as follows:

Both systems display a second-order QPT at a certain critical value
$U_c$ of the atom-atom interaction. In both cases, for $U<U_c$ the
ground state is degenerate due to the breaking of the left-right
symmetry in the atomic system, giving rise to a finite population
imbalance.  On the contrary, for $U>U_c$ the ground state is not
degenerate and the imbalance is zero. This critical value of the
control parameter was obtained by means of a mean-field calculation,
and it was numerically tested by means of exact diagonalizations and
finite-size scaling using the gap between the ground state and the
first excited state as a signature of the transition. We have also
related the population imbalance with the square root of the
Fisher information, and tested the mean-field with exact
diagonalizations.

Beyond mean-field techniques that take into account the quantum
fluctuations have been used to describe the low energy spectrum. A
comparison with exact diagonalizations shows the goodness of these
theoretical results. In particular, it is worth mentioning that the
theory successfully accounts for the crossings between levels
corresponding to opposite parities, which take place when the degree
of freedom of the diatomic molecule is added. This formalism greatly
improves the description of the critical point, showing that the
second-order QPT is strongly influenced by correlations and
fluctuations that are beyond the mean-field approximation.

Both systems also display ESQPTs at certain critical energies
depending on the value of the interaction $U$. Contrary to the ground
state QPT, the nature of these transitions differs in both models. The
two-site Bose-Hubbard Hamiltonian is characterized by a $\lambda$
divergence in the density of states. This behavior in the energy
spectrum is comparable to the Lipkin and the Tavis-Cummings models,
both having a single semiclassical degree of freedom. On the contrary,
the density of states for the case in which the atoms interact with a
diatomic molecule, shows no divergence. This behavior is qualitatively
similar to the Dicke model, in which the divergence arises in the
derivative of the density of states.

Finally, we can infer a correlation between the ESQPT and
  the onset of chaos in the systems with atoms and molecules. A similar
  conclusion has been obtained for the Dicke model \cite{Pedro:11b},
  though it seems that the transition between the integrable and the
  chaotic regime does not coincide with the critical energy at which
  the ESQPT takes place \cite{Hirsch:14}. Taking into
  account the strong numerical limitations for the atom-molecule
  system, we conjecture that quantum chaos only appears in the
  non-degenerate region of the spectrum. Whether this
  is a generic mechanism, linking chaos and critical phenomena in the
  energy spectrum of systems with more than one semiclassical degree
  of freedom remains still an open question.

\acknowledgments

This work is has been partially supported by the Spanish MINECO grants
FIS2012-35316, FIS2012-34479, and 
FIS2011-28738-c02-01, by Junta
de Andaluc\'{\i}a under group number FQM-160 and Project P11-FQM-7632, 
and by the Consolider-Ingenio 2010 Programme CPAN (CSD2007-00042).



\end{document}